\documentclass[universe,article,accept,pdftex,moreauthors]{Definitions/mdpi}
\graphicspath{{./figures/}}
\firstpage{1}
\pubvolume{9}
\issuenum{4}
\articlenumber{175}
\pubyear{2023}
\copyrightyear{2023}
\externaleditor{{Academic Editor: Luigi Foschini} 
 }
\datereceived{17 March 2023 }
\daterevised{31 March 2023 } 
\dateaccepted{2 April 2023}
\datepublished{3 April 2023 }
\hreflink{https://doi.org/10.3390/\linebreak universe9040175} 

\usepackage{amssymb}

\Title{{Estimate} of SMBH Spin for Narrow-Line Seyfert 1 Galaxies\\ \small{\it{Accepted for publication in Universe}}}

\TitleCitation{Estimate of SMBH Spin for Narrow-Line Seyfert 1 Galaxies}

\Author{{Mikhail Piotrovich} 
 *\orcidA{}, Stanislava Buliga \orcidB{} and Tinatin Natsvlishvili \orcidC{}}

\AuthorNames{Mikhail Piotrovich, Stanislava Buliga and Tinatin Natsvlishvili}

\AuthorCitation{Piotrovich, M.; Buliga, S.; Natsvlishvili, T.}

\address[1]{%
Central Astronomical Observatory {at} 
 Pulkovo RAS, 196140 Saint-Petersburg, Russia; aynim@yandex.ru (S.B.); tinatingao@mail.ru (T.N.)}

\corres{\hangafter=1 \hangindent=1.05em \hspace{-0.82em}
Correspondence: mpiotrovich@mail.ru}

\abstract{We estimated {{the}
} spin values of { the} supermassive black holes (SMBHs) of { the} active galactic nuclei (AGN) for { a} large set of Narrow Line Seyfert 1 (NLS1) galaxies { assuming the inclination angle between the line of sight and the axis of the accretion disk to be approximately 45 degrees}. We found that for these objects the spin values are on average less than for the Seyfert 1 galaxies that we studied previously. In addition, we found that the dependencies of { the} spin on { the} bolometric luminosity and { the} SMBH mass are two to three times stronger that for Seyfert 1 galaxies, which could mean that at early stages of evolution NLS1 galaxies either have { a} low accretion rate or chaotic accretion, while at later stages they have standard disk accretion, which very effectively increases the spin value.}


\begin{document}

\section{Introduction}

{ There are many types of active galaxies with Active Galactic Nuclei (AGNs), such as Seyfert galaxies, quasars, BL Lac objects, and radio galaxies. Among these types of galaxy, there are two types that are determined by the properties of their emission lines,} namely, Seyfert 1 ({Sy1}
) and Seyfert 2 ({Sy2}) galaxies~\citep{netzer15}. {Sy1} galaxies exhibit both broad allowed emission lines from the broad line region (BLR) with a width of several thousand~km/s, and narrow forbidden emission lines from the narrow line region (NLR) with a width of several hundreds of~km/s. {Sy2} galaxies are characterized by narrow allowed and forbidden lines in their emission spectra~\citep{robson96}. According to the \citet{antonucci93} model, both types of galaxies, Sy1 and Sy2, have a similar internal structure, and the differences in their spectra are mainly due to orientation effects. Although the differences between Sy1 and Sy2 galaxies are well defined, galaxies with narrow resolved emission lines, similar to Sy2 galaxies but having all the spectral properties of Sy1 sources, have been found as well. These galaxies have been classified as narrow-line Sy1 (NLS1) galaxies~\citep{osterbrock85}. Galaxies of this type are characterized by the following: (1) full width at half maximum (FWHM) of the broad line H$\beta$~<~2000~km/s~\citep{goodrich89}; (2) weak [OIII] emission lines relative to H$\beta$ with [OIII] $\lambda 5007$/H$\beta$~<~3~\citep{osterbrock85,leighly99}; (3) strong emission lines of FeII relative to H$\beta$ in the ultraviolet and optical regions of the spectrum~\citep{mathur00}; (4) a strong excess of soft X-rays and high amplitude of rapid X-ray variability~\citep{boller96,leighly99}; (5)~strong infrared emissions, indicating active star formation~\citep{moran96}.

{ The} SMBHs in AGNs are characterized by two main parameters, namely, the mass and the spin (dimensionless angular momentum). The spin is very important because, according to modern concepts, the radiative efficiency of the accretion disk (among other things) strongly depends on the value of the spin~\citep{bardeen72,novikov73,krolik07,krolik07b}.

In our previous works, we mainly explored AGNs in Seyfert 1-type galaxies; thus, in this work, we decided to study AGNs in NLS1 and to compare these two types of AGNs.

\section{Examination of Initial Data}

We took initial data from \citet{zhou06cat}. This catalogue consists of 2011 NLS1 type objects; of { these}, 2005 have all the necessary data for our calculations. These include $L_\text{5100}$---luminosity at 5100 \AA\, $FWHM$(H$\beta$)---full width at half maximum of { the} H$\beta$ spectral line (which determines the rotation speed of an accretion disk of AGN), { and $z$---the cosmological~redshift}.

\begin{figure}[H]
  \includegraphics[bb= 75 20 735 535, clip, width=0.625\textwidth]{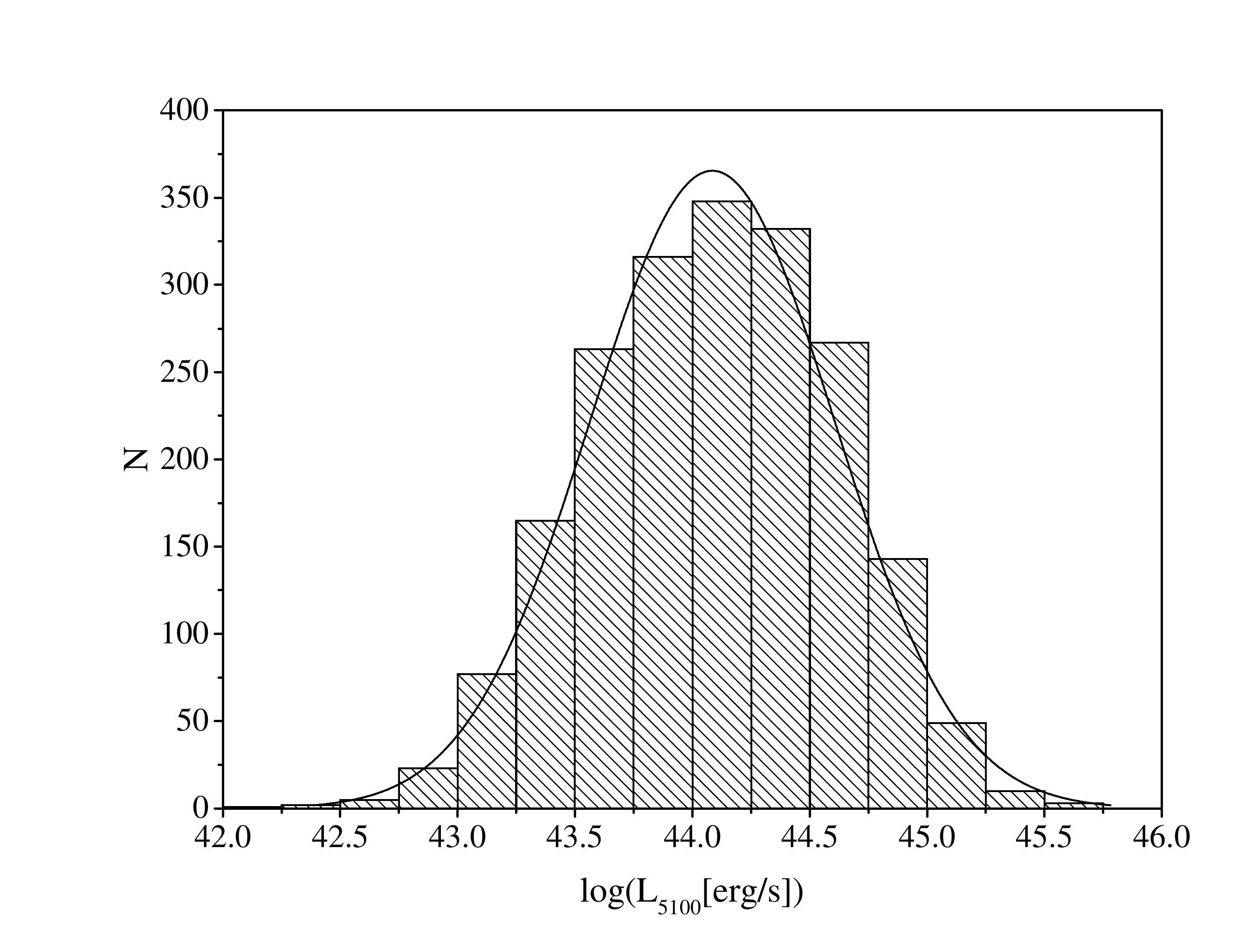}
  \caption{The distribution of the luminosity at 5100 \AA\, $L_\text{5100}$ {(with a fitted log-normal distribution curve)} from the initial set.}
  \label{fig01}
\end{figure}

\begin{figure}[H]
  \includegraphics[bb= 75 20 735 535, clip, width=0.625\textwidth]{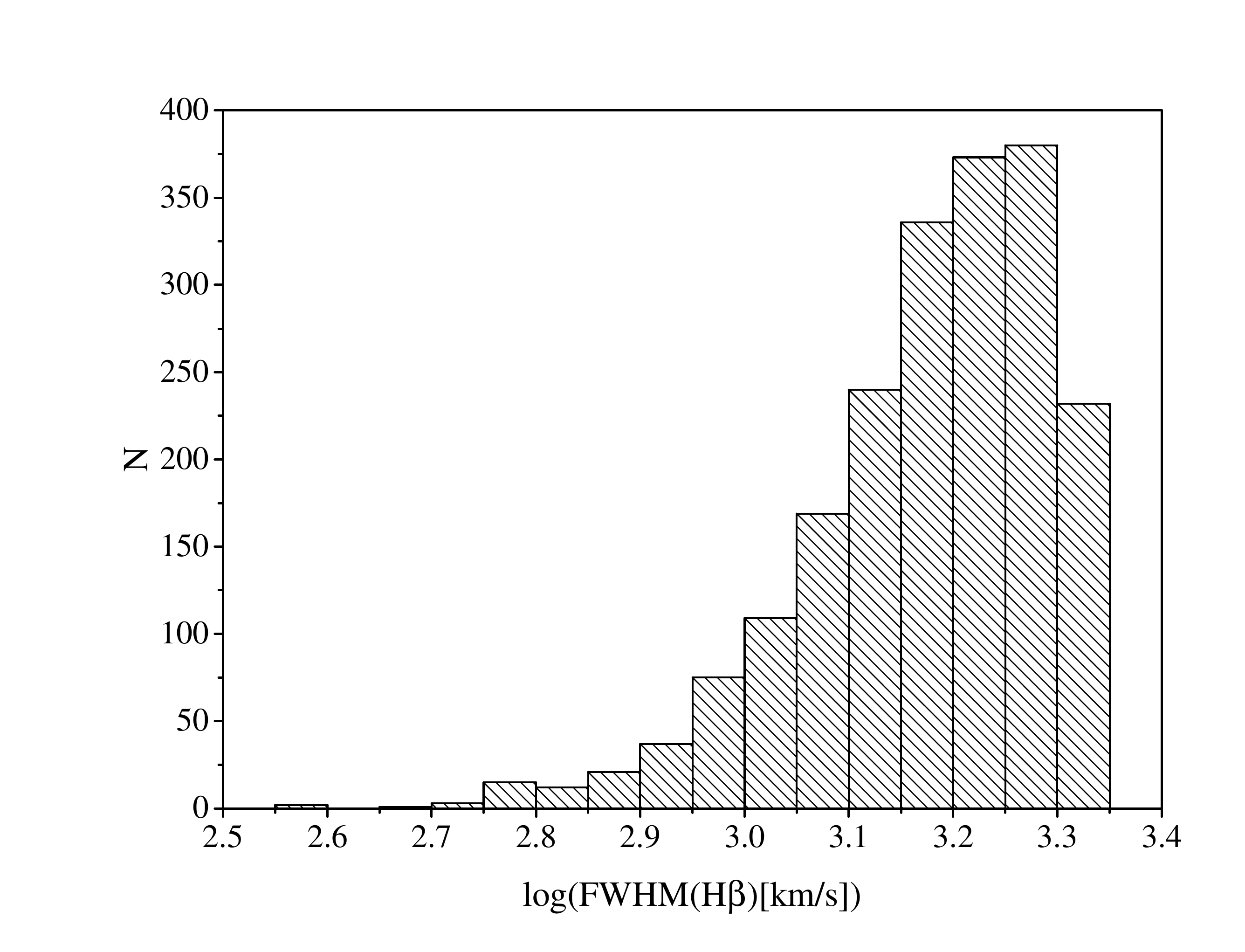}
  \caption{The distribution of $FWHM$(H$\beta$) from the initial set.}
  \label{fig02}
\end{figure}

\begin{figure}[H]
  \includegraphics[bb= 75 20 735 535, clip, width=0.65\textwidth]{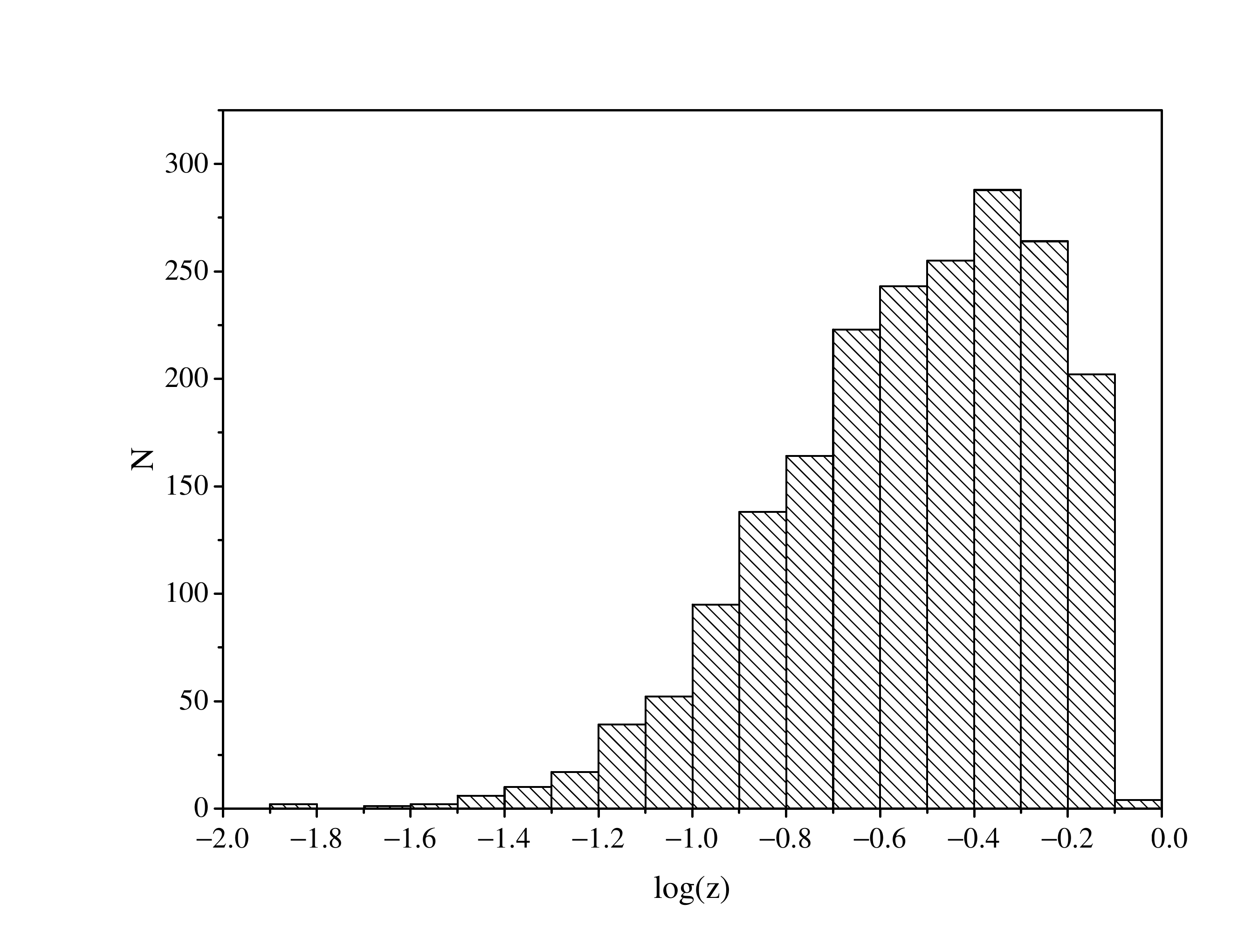}
  \caption{{The} 
 distribution of the cosmological redshift $z$ from the initial set.}
  \label{fig03}
\end{figure}

\begin{figure}[H]
  \includegraphics[bb= 70 40 680 535, clip, width=0.65\textwidth]{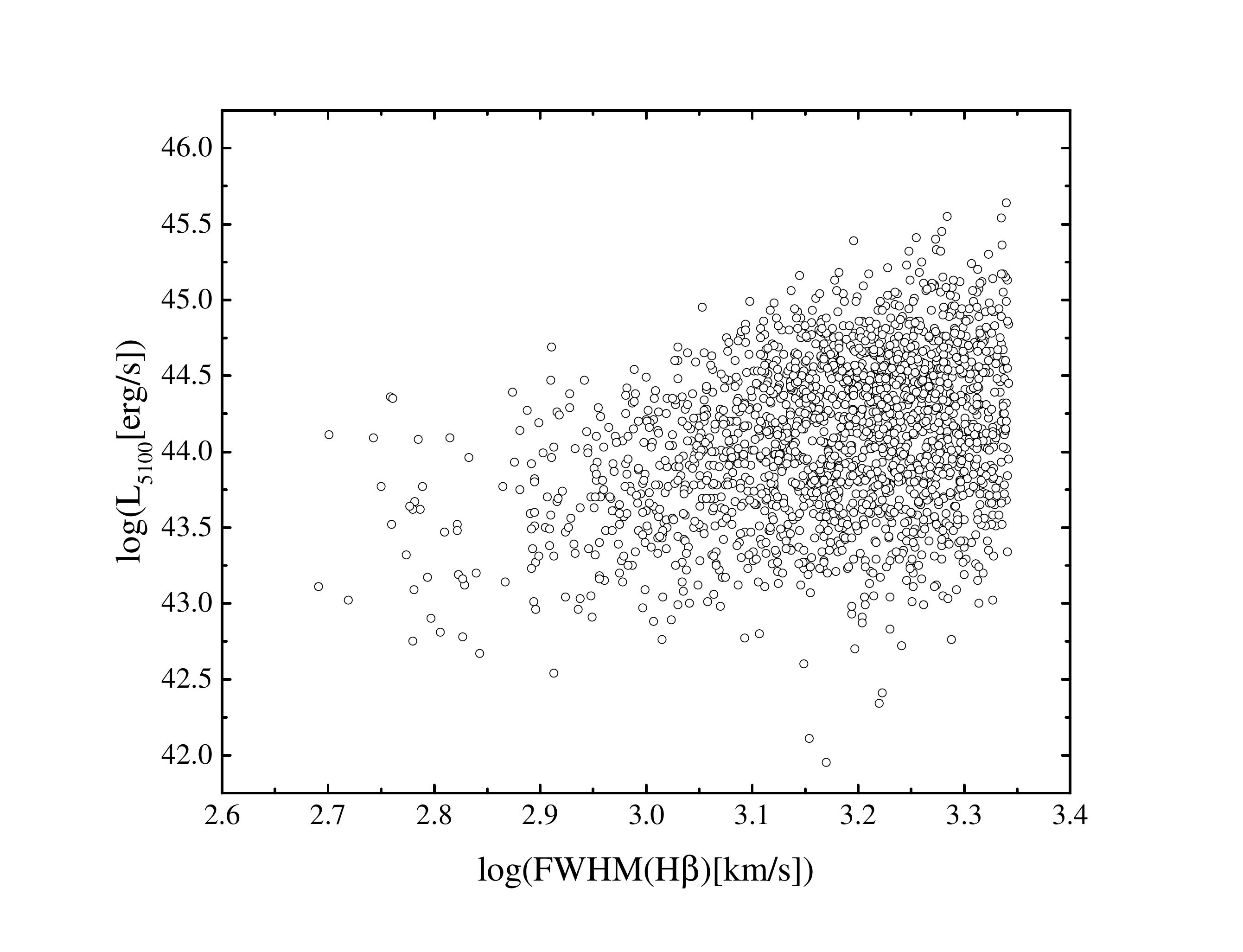}
  \caption{The dependence of $FWHM$(H$\beta$) on the luminosity at 5100 \AA\, $L_\text{5100}$ from the initial set.}
  \label{fig04}
\end{figure}

First, we examine the data from { the} catalogue. Figures~\ref{fig01}--\ref{fig03} show the histograms with the distributions of $L_\text{5100}$, $FWHM$(H$\beta$) and the cosmological redshift $z$. We can see that { the} 5100 \AA\, that is, the luminosity, has a { log-normal} distribution with its peak at $\log(L_\text{5100}\text{[erg/s]}) \approx 44$. Concerning $FWHM$(H$\beta$), it can be seen that the right-hand side of the distribution ends quite abruptly at $\log$($FWHM$(H$\beta$)[km/s]) $\approx 3.4$. This can be explained by the fact that AGNs in NLS1-type objects are characterized by a lower rotation speed of the accretion disk compared to, for example, Seyfert 1 galaxies, as well as by the method used to process { the} data during the creation of the catalogue~\citep{zhou06}.{ The} cosmological redshift distribution may be caused by the spatial distribution of objects (for close objects) and the selection effect (for distant objects). Figure~\ref{fig04} shows { the} dependence of { the} luminosity at 5100 \AA\, $L_\text{5100}$ on $FWHM$(H$\beta$). { The Spearman correlation coefficient for this parameter is 0.28, and the correlation is significant at the 0.05 level. Thus, we can only note a weak correlation between the parameters, which is as expected.}
\vspace{-4pt}
\vspace{-8pt}

\vspace{-12pt}
\vspace{-12pt}

\section{Estimation of Spin Values}

The spin (dimensionless angular momentum) of an SMBH is defined as $a = c J / G M_\text{BH}^2$, where $J$ is the angular momentum, $M_\text{BH}$ is the mass of the black hole, and $c$ is the speed of light. We can obtain the spin via the radiative efficiency $\varepsilon = L_\text{bol} / \dot{M} c^2$ (where $L_\text{bol}$ is the bolometric luminosity of AGN and $\dot{M}$ is the accretion rate), which depends strongly on the spin of the SMBH~\citep{bardeen72,novikov73,krolik07,krolik07b}.

There are several models~\citep{davis11,raimundo11,du14,trakhtenbrot14,lawther17} connecting the radiative efficiency with { the} physical parameters of AGNs, which can be obtained from observations. In our previous work~\citep{piotrovich22}, we concluded that { the} model from \citet{du14} provides the most consistent results; therefore, we decided to use it in this work.

According to \citet{du14}:
\begin{equation}
 \varepsilon \left( a \right) =  0.105 \left(\frac{L_\text{bol}}{10^{46}\,\text{erg/s}}\right) \left(\frac{L_\text{5100}}{10^{45}\,\text{erg/s}} \right)^{-1.5} M_8 \mu^{1.5},
 \label{epsilon}
\end{equation}

\noindent where $M_8 = M_\text{BH} / (10^8 M_{\odot})$,  $\mu = \cos{i}$, and $i$ is { the} angle between the line of sight and the axis of the accretion disk.{ Because the angles for most objects are unknown, and because we have no reason to assume the presence of any preferred direction in the orientation of galaxies, the generally accepted method is to assume some average angle. As there is insufficient statistical data on the preferred angle for NLS1, we assume $i = 45^{\circ}$, $\mu = 0.7$. Although this method is not perfect, there are a number of arguments in its favor; for example, existing angle measurements (including those made with the participation of the authors of the present work~\citep{afanasiev18}) show that for most objects of the Seyfert 1 type (including NLS1) they usually range from 20 to 60 degrees, and for a noticeable number of objects the angle is close to 45 degrees (see, for example, \citet{marin16,afanasiev18}), which reduces the possible size of errors.} Furthermore, in our calculations we use the Eddington ratio $l_\text{E} = L_\text{bol} / L_\text{Edd}$, where $L_\text{Edd} = 1.3 \times 10^{38} M_\text{BH} / M_\odot$ is the Eddington luminosity.

In order to obtain { the} bolometric luminosity $L_\text{bol}$ from $L_\text{5100}$, we need to use bolometric correction. Various authors provide different bolometric corrections that can differ by up to two to three times~\citep{richards06, hopkins07, cheng19, netzer19, duras20}. We tested several different methods of bolometric correction, and for consistency decided to use the same method we used in \citet{piotrovich22}, namely, { the} approach from \citet{richards06}: $L_\text{bol} = L_\text{5100} \times 10.3$.

We obtained { the} masses of { the} SMBHs using { the} method from \citet{vestergaard06}:
\begin{equation}
    \log(M_\text{BH}) \approx \log\left(\left[\frac{FWHM(H\beta)}{1000\, \text{km/s}}\right]^2 \left[\frac{L_\text{5100}}{10^{44}\, \text{erg/s}}\right]^{0.5}\right) + 6.91.
    \label{mass}
\end{equation}

{ The} radiative efficiency for this type of object must satisfy the condition $0.039~<~\varepsilon(a) < 0.324$~\citep{thorne74}. In addition, because { the} method from~\citep{du14} uses { the} Shakura--Sunyaev accretion disk model~\citep{shakura73}, the Eddington ratio must be in the range $0.01 \leq l_\text{E} \leq 0.3$~\citep{netzer14}. Thus, from 2005 initial objects, we obtained 474 objects satisfying these conditions. The spin $a$ was determined numerically using the expression from \citet{bardeen72}:
\begin{equation}
 \varepsilon(a) = 1 - \frac{R_\text{ISCO}^{3/2} - 2 R_\text{ISCO}^{1/2} + |a|}{R_\text{ISCO}^{3/4}(R_\text{ISCO}^{3/2} - 3 R_\text{ISCO}^{1/2} + 2 |a|)^{1/2}},
 \label{epsilon_R_ISCO}
\end{equation}

\noindent where $R_\text{ISCO}$ is the radius of the innermost stable circular orbit of { the} SMBH and
\begin{equation}
 \begin{array}{l}
  R_\text{ISCO}(a) = 3 + Z_2 \pm ((3 - Z_1)(3 + Z_1 + 2 Z_2))^{1/2},\\
  Z_1 = 1 + (1 - a^2)^{1/3}((1 + a)^{1/3} + (1 - a)^{1/3}),\\
  Z_2 = (3 a^2 + Z_1^2)^{1/2}.
 \end{array}
 \label{R_ISCO}
\end{equation}

 In the expression for $R_\text{ISCO}(a)$, the sign ''{$-$}
'' is used to indicate { the} prograde ($a \geq 0$), while the sign ''+'' indicates { the} retrograde rotation ($a~<~0$).

The results of our calculations are presented in Tables~\ref{table1_1}--\ref{table1_8}. We must emphasize that the obtained spin values are of course not exact, and can only be considered as estimates { intended for statistical analysis. Nevertheless, in the absence of other spin data, they can be used as a first approximation.}

\section{Analysis of Objects Using Estimated Spin Values}

First, we consider { the} statistical properties of { the} objects using the estimated spin values (the new set) as compared to the initial set.

\begin{figure}[H]
  \includegraphics[bb= 75 20 735 535, clip, width=0.65\textwidth]{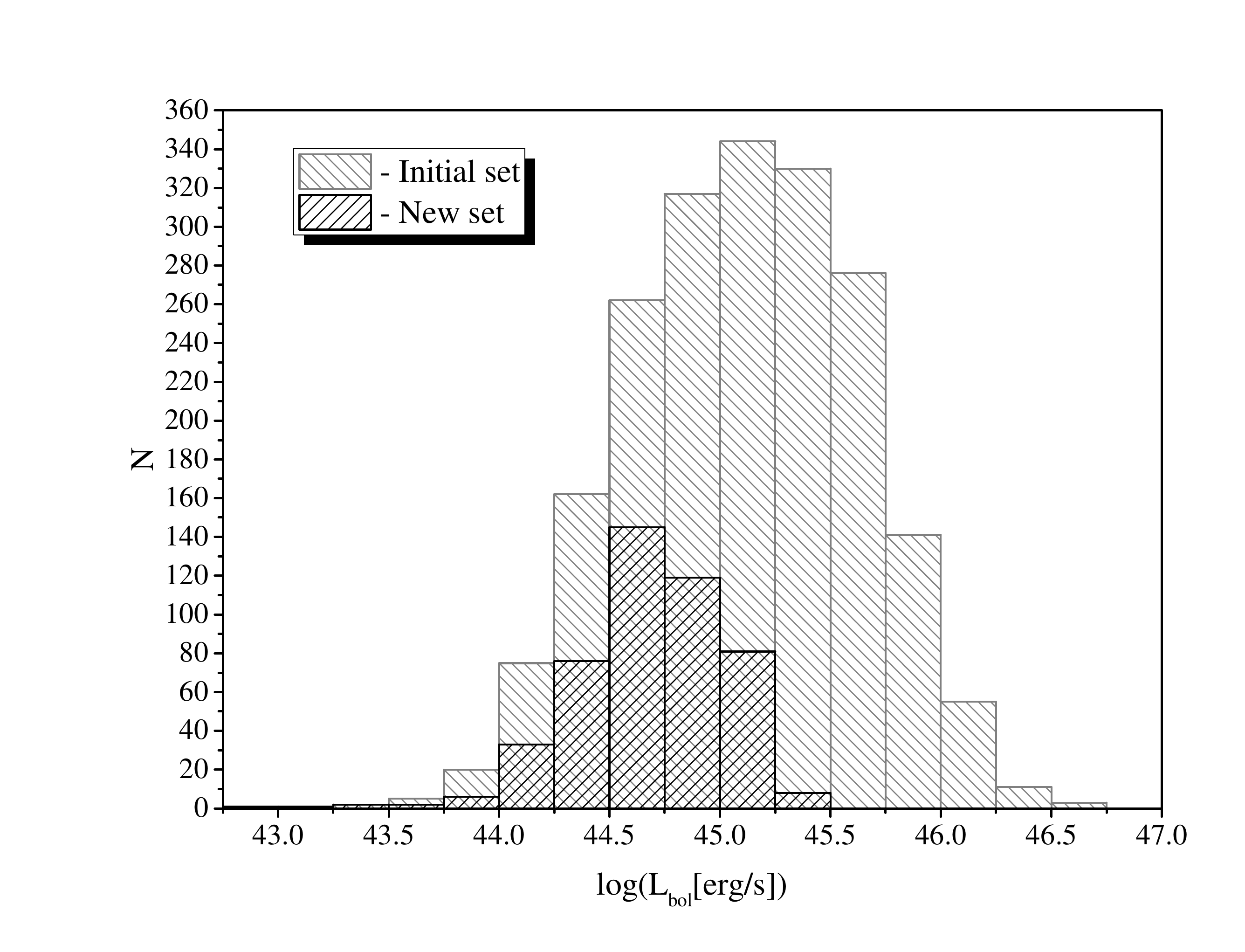}
  \caption{The distribution of the bolometric luminosity for the initial and new sets.}
  \label{fig05}
\end{figure}

Figure~\ref{fig05} shows the distribution of { the} bolometric luminosity for both sets. It can be seen that both the new set and the initial one have normal distributions; however, in the new set the peak is shifted to the left by an order of magnitude. This can be explained by the fact that we estimated spins only for { those} objects with Eddington ratios $l_\text{E}~<~0.3$.

\begin{figure}[H]
  \includegraphics[bb= 75 20 735 535, clip, width=0.65\textwidth]{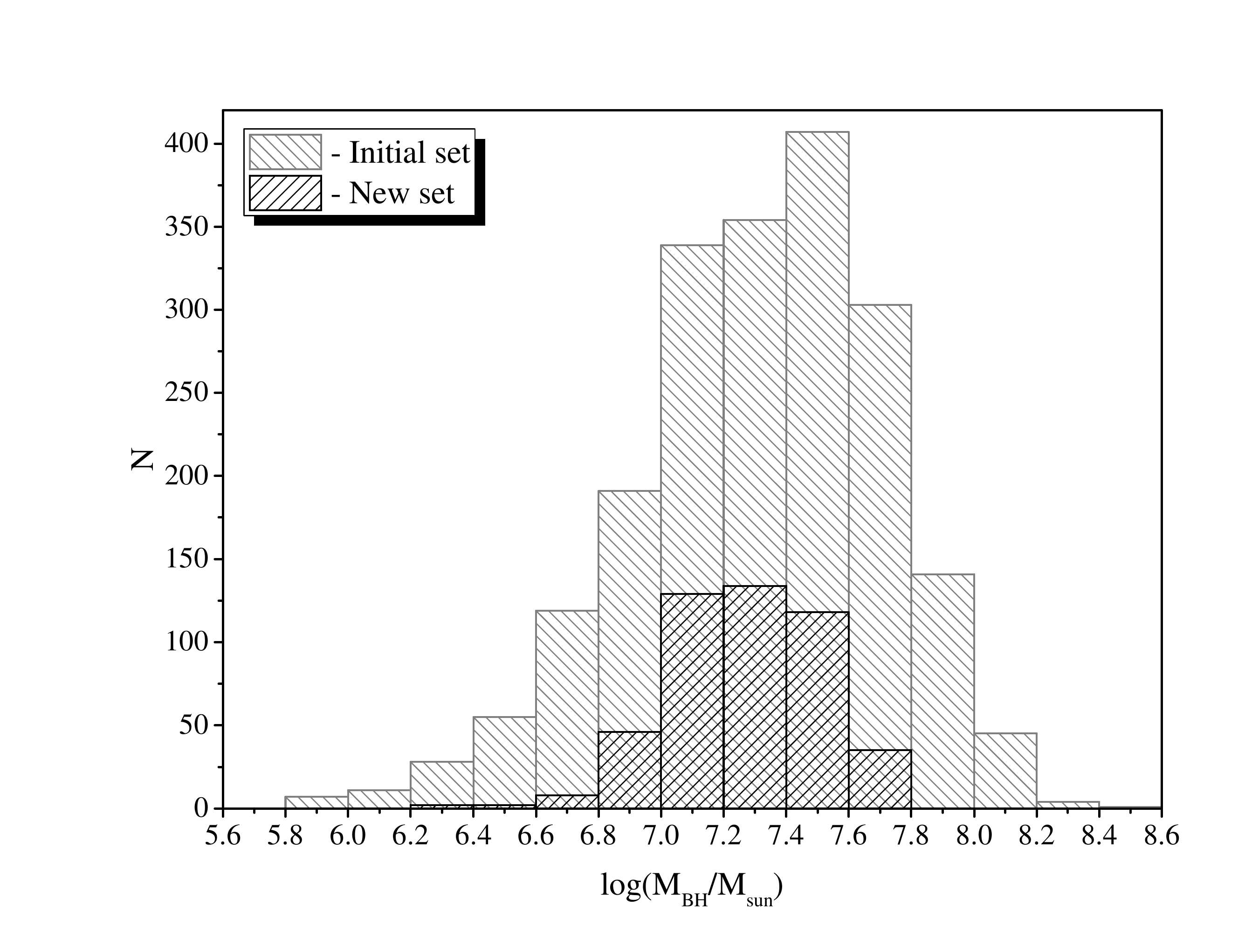}
  \caption{The distribution of the SMBH mass for the initial and new sets.}
  \label{fig06}
\end{figure}

Figure~\ref{fig06} demonstrates the distribution of { the} SMBH mass for both sets. In general, the distributions have a similar form.

\begin{figure}[H]
  \includegraphics[bb= 75 20 735 535, clip, width=0.65\textwidth]{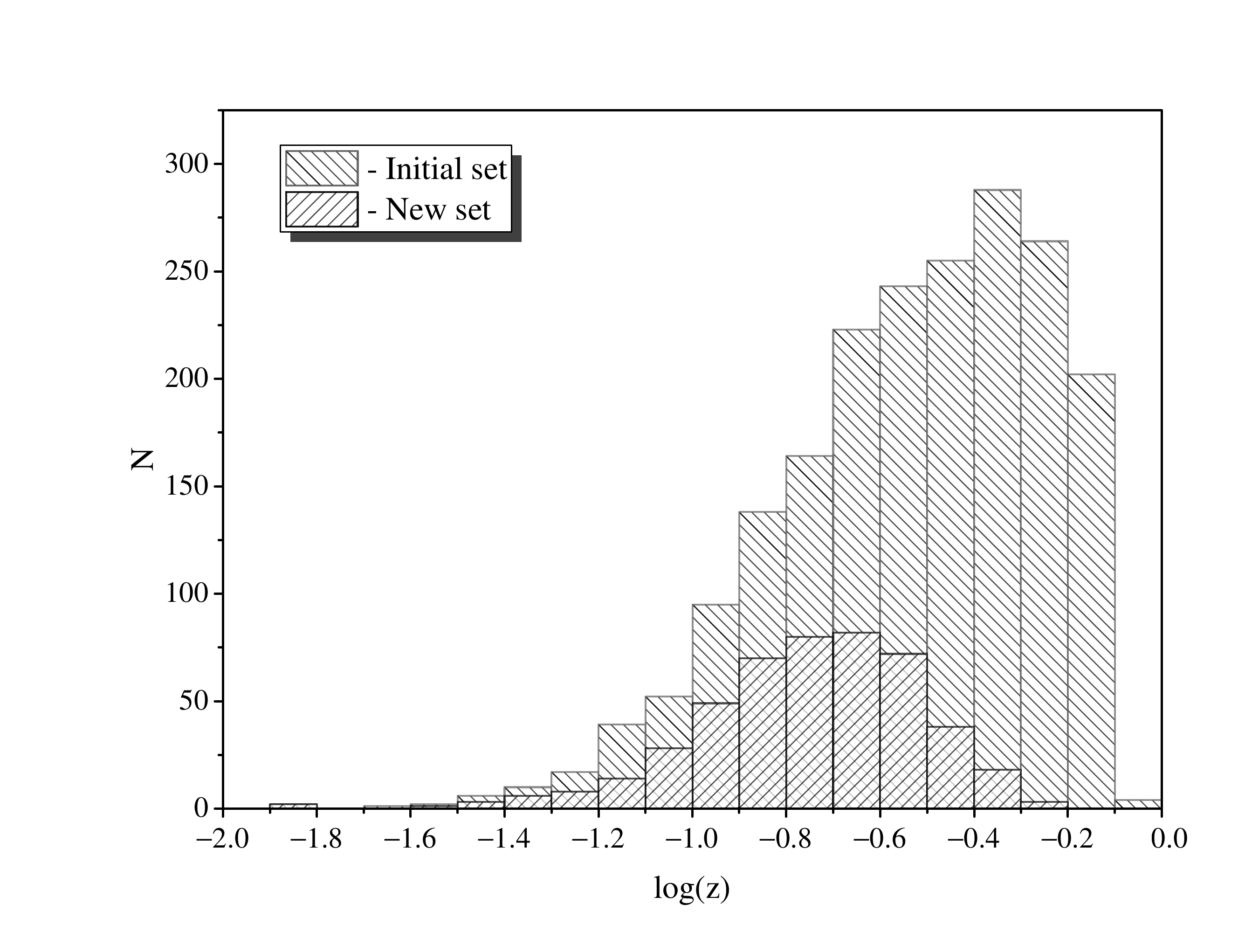}
  \caption{{The} 
 distribution of the cosmological redshift for the initial and new sets.}
  \label{fig07}
\end{figure}

Figure~\ref{fig07} shows the distribution of cosmological redshift for both sets. While this distribution again has similar form, its peak is shifted towards closer objects. This occurs because the new set consists of { predominantly} fainter objects that we are usually only able to { detect} at closer distances (selection effect).

\begin{figure}[H]
  \includegraphics[bb= 75 20 735 535, clip, width=0.65\textwidth]{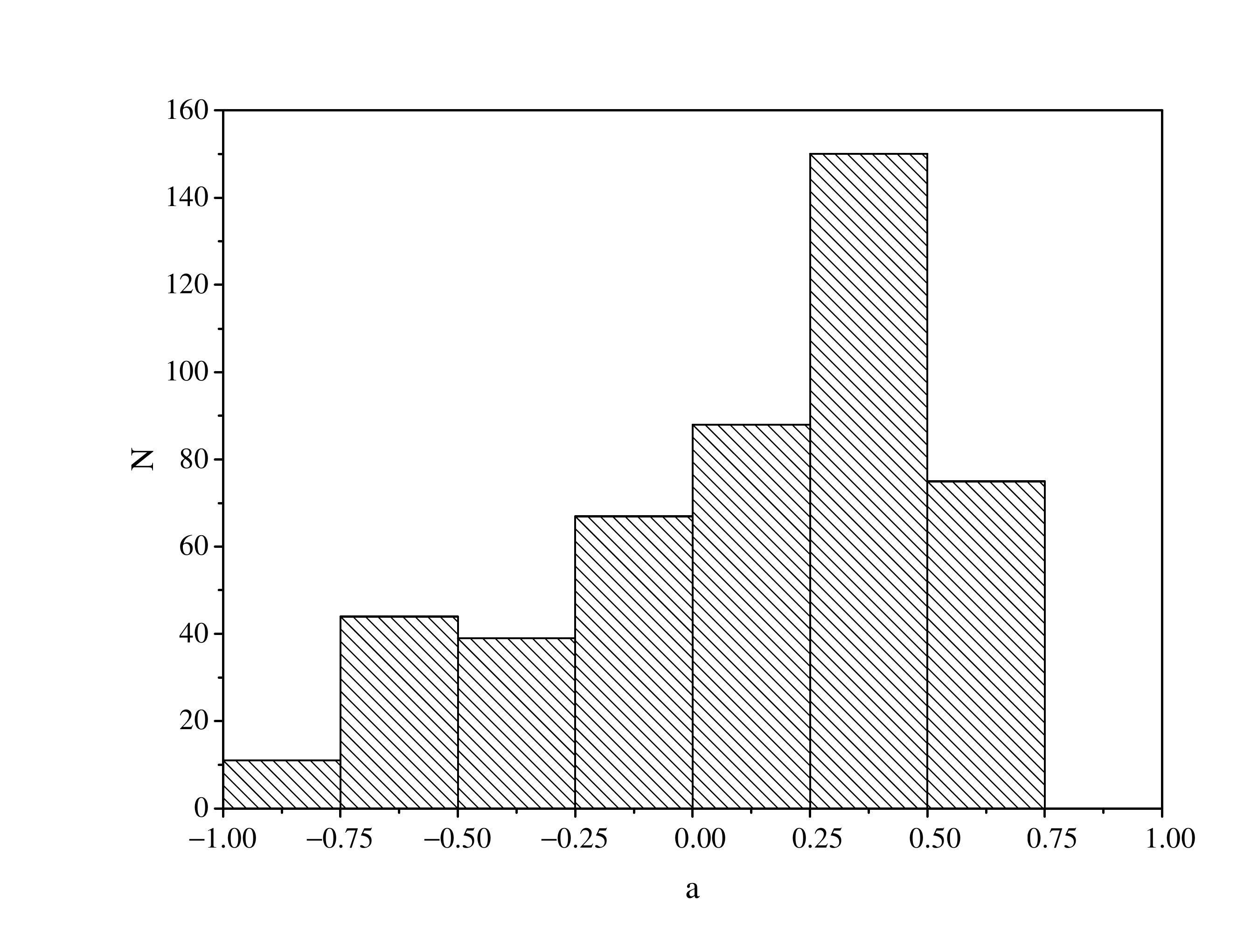}
  \caption{{The} 
 distribution of the estimated spin values $a$.}
  \label{fig08}
\end{figure}

\begin{figure}[H]
  \includegraphics[bb= 80 20 685 535, clip, width=0.6\textwidth]{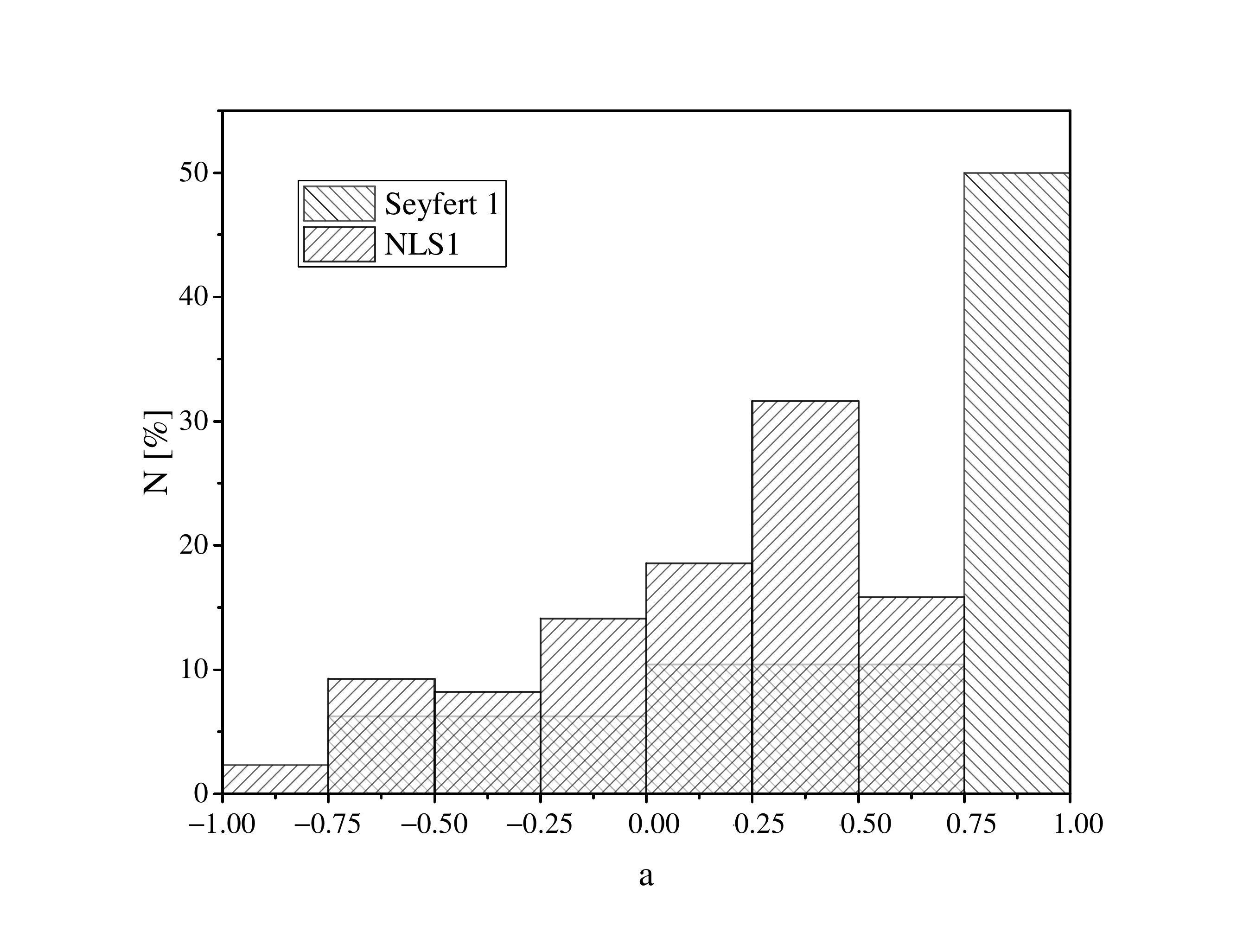}
  \caption{{The} 
 normalized distributions of the estimated spins for Seyfert 1~\citep{piotrovich22} and NLS1 (this~work).}
  \label{fig08_1}
\end{figure}

Figure~\ref{fig08} shows the distribution of the estimated spin values for { the} 474 objects. The distribution has a pronounced peak at $0.25~<~a~<~0.5$ and terminates at $a > 0.75$. This is very different from the typical distribution for Seyfert 1-type objects (see Figure~\ref{fig08_1}), for which the distribution usually has its peak at $0.75~<~a~<~1.0$ and up to 50\% of objects have spin values $a > 0.75$~\citep{trakhtenbrot14,afanasiev18,piotrovich22}. This result is generally consistent with the results of \citet{liu15} obtained via X-ray observations.

It is interesting to compare our results with { the} distribution of spins estimated in \citet{chen23} (see Figure 6 in their work) for various types of active galaxies. It can be observed that our distribution of spin values for NLS1 looks very similar to their distribution for radio galaxies, which may indicate that these two types of objects are closely related~\citep{yuan08,berton18}. In addition, it can be seen that our spin distribution for Seyfert 1-type galaxies resembles their distribution for flat-spectrum radio quasars (FSRQ), which in turn could mean that Seyfert 1 galaxies and FSRQs are related { (for example, it may mean that these are objects of the same type observed from different directions)}.

\begin{figure}[H]
  \includegraphics[bb= 70 35 680 535, clip, width=0.6\textwidth]{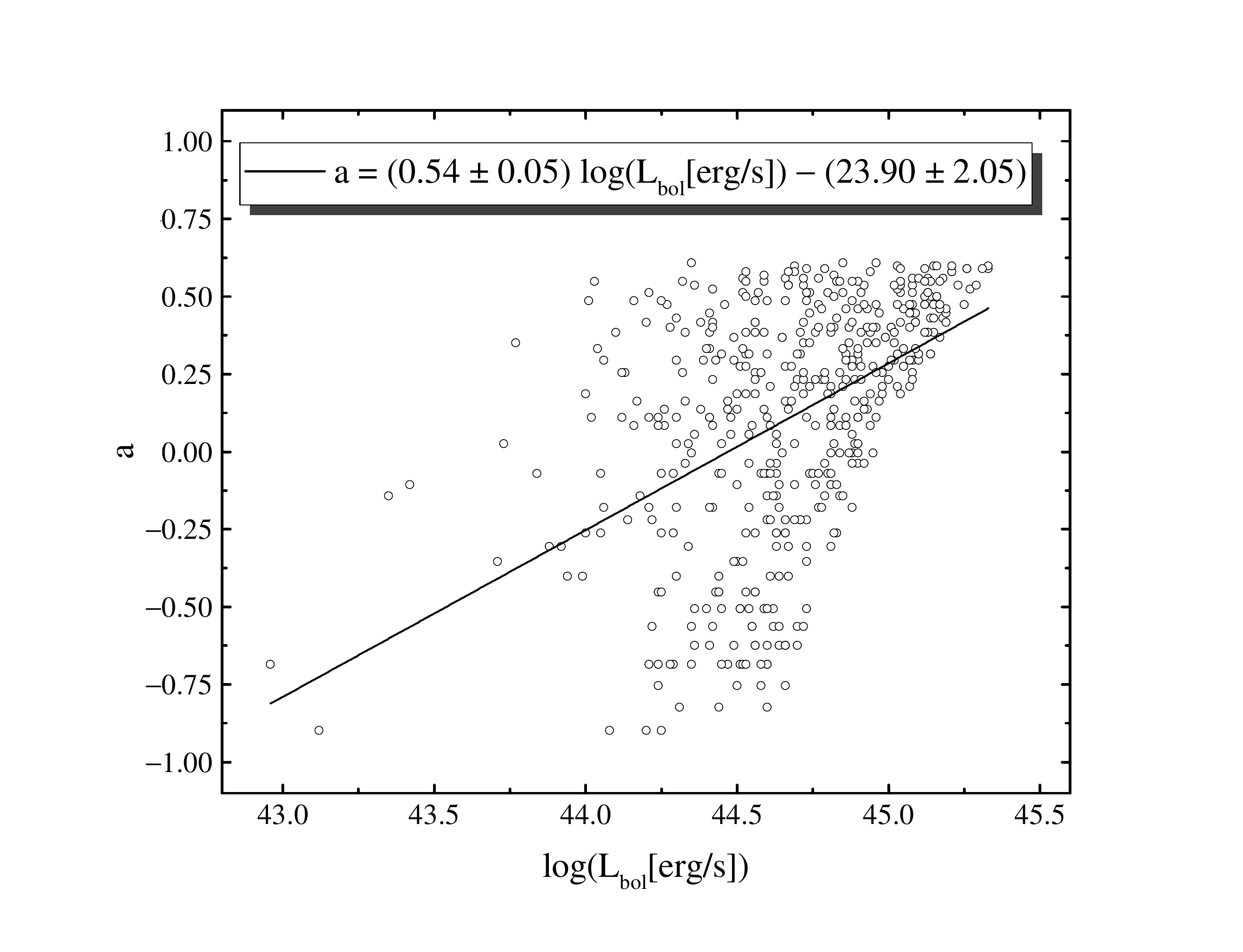}
  \caption{{The} 
 dependence of the estimated spin values $a$ on the bolometric luminosity $L_\text{bol}$.}
  \label{fig09}
\end{figure}

Figure~\ref{fig09} shows the dependence of the estimated spin values $a$ on the bolometric luminosity $L_\text{bol}$. { The} linear fitting provides us with the following expression:
\begin{equation}
    a = (0.54 \pm 0.05) \log(L_\text{bol}\text{[erg/s]}) - (23.90 \pm 2.05).
    \label{a_L}
\end{equation}

 In our previous work in \citet{piotrovich22}, the following was obtained for Seyfert 1 galaxies: $a = (0.25 \pm 0.07) \log{L_\text{bol}\text{[erg/s]}} - (10.59 \pm 3.21)$. { Thus,} we can conclude that { the} dependence of the spin on the bolometric luminosity for the NLS1 type is stronger than for Seyfert 1 galaxies.

\begin{figure}[H]
  \includegraphics[bb= 70 35 680 535, clip, width=0.6\textwidth]{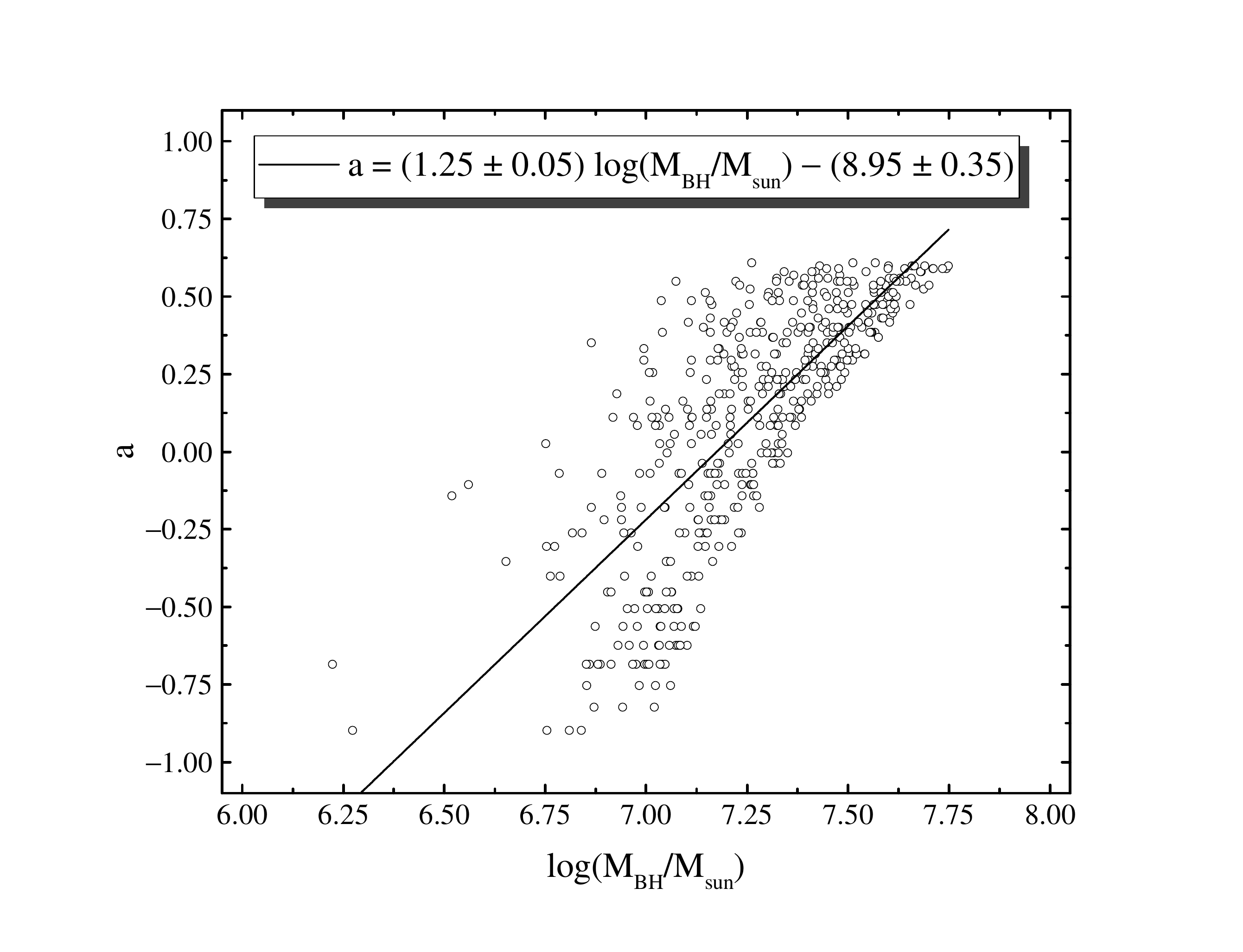}
  \caption{{The} 
 dependence of the estimated spin values $a$ on the SMBHs masses $M_\text{BH}$.}
  \label{fig10}
\end{figure}

\fussy Figure~\ref{fig10} shows the dependence of the estimated spin values $a$ on the mass of { the} SMBH $M_\text{BH}$. { The} linear fitting provides us with the following expression:
\begin{equation}
    a = (1.25 \pm 0.05) \log(M_\text{BH}/M_\odot) - (8.95 \pm 0.35).
    \label{a_M}
\end{equation}

 In \citet{piotrovich22}, we obtained the following for Seyfert 1 galaxies: $a = (0.46 \pm 0.09) \log{M_\text{BH}/M_\odot} - (3.00 \pm 0.71)$. { Thus,} we can conclude that the dependence of the spin on { the} SMBH mass is stronger for the NLS1 type than for Seyfert 1 galaxies.

Together with the low average spin, this could mean that at early stages of evolution NLS1 either have low accretion rates or chaotic accretion, while at later stages (which we are studying in this work) they have standard disk accretion, which very effectively increases the spin value.
\vspace{-4pt}
\vspace{-8pt}
\vspace{-8pt}
\vspace{-8pt}
\vspace{-8pt}
\vspace{-12pt}

\vspace{-8pt}

\section{Conclusions}

In this work, we have estimated spin values of { the} SMBHs in AGNs for 474 NLS1-type galaxies, { assuming the inclination angle between the line of sight and the axis of the accretion disk $i \approx 45^{\circ}$}.{ The} distribution of the estimated spin values differs significantly from { the} distribution of { the} spins for Seyfert 1-type galaxies. On average, { the} spin values are smaller.{ The} distribution has a peak at $0.25<a<0.5$, and there are no objects with spins $a > 0.75$. This is generally consistent with { the} results of \mbox{\citet{liu15}}. Our distribution of spin values for NLS1 looks very similar to { the} distribution for radio galaxies from \citet{chen23}, which may indicate that these two types of objects are closely related~\citep{yuan08,berton18}. In addition, it can be seen that our spin distribution for Seyfert 1-type galaxies in our previous works resembles {the} distribution from \mbox{\citet{chen23}} for flat-spectrum radio quasars (FSRQ), which in turn could mean that Seyfert 1 galaxies and FSRQs are related {(for example, it may mean that these are objects of the same type observed from different directions)}. The dependencies of spin on { the} bolometric luminosity (Equation~\eqref{a_L}) and SMBH mass (Equation~\eqref{a_M}) are quite different from { the} Seyfert 1 case. Specifically, { the} dependencies of the spin on these parameters is two to three times stronger, which could mean that in the early stages of evolution the NLS1 type either have a low accretion rate or chaotic accretion, while at later stages they have standard disk accretion, which very effectively increases the spin value.

\vspace{6pt}
\authorcontributions{Conceptualization, M.P.; methodology, M.P.; software, M.P.; validation, S.B. and T.N.; formal analysis, M.P.; investigation, M.P., S.B. and T.N.; resources, M.P., S.B. and T.N.; data curation, S.B. and T.N.; writing---original draft preparation, M.P.; writing---review and editing, M.P.; visualization, M.P.; supervision, M.P.; project administration, M.P. All authors have read and agreed to the published version of the manuscript.}

\funding{This research received no external funding.}

\institutionalreview{{Not applicable.} 
}

\informedconsent{{Not applicable.} 
}

\dataavailability{The data underlying this article are available in the article.}

\conflictsofinterest{The authors declare no conflict of interest.}

\setcounter{section}{0}
\appendixtitles{no}
\appendix
\section{}

\vspace{-6pt}
\begin{table}[H]
\caption{{All} 
 objects with their estimated spin values. Part 1 of 8. Columns indicate: (1) object name, (2)~cosmological redshift, (3) bolometric luminosity, (4) mass of SMBH, and (5) estimated spin value.}
\footnotesize
\begin{tabularx}{\textwidth}{m{4cm}<{\raggedright}m{1.3cm}<{\centering}m{2.5cm}<{\centering}m{1cm}<{\centering}m{1.3cm}<{\centering}m{1.2cm}<{\centering}}
\toprule
\textbf{Object}	& \textbf{\boldmath$z$} & \textbf{\boldmath$\log(L_\text{bol}\text{[erg/s]})$} & \textbf{\boldmath$l_\text{E}$} & \textbf{\boldmath$M_\text{BH}/M_\odot$} & \textbf{\boldmath$a$}\\
\midrule
SDSS J000154.27{+}
000732.4 & 0.139595 & 44.32 & 0.126 & 7.11 & 0.254 \\
SDSS J001010.03+005126.6 & 0.387000 & 45.18 & 0.293 & 7.60 & 0.430 \\
SDSS J001137.24+144201.4 & 0.131834 & 44.74 & 0.165 & 7.41 & 0.512 \\
SDSS J001630.43{-}
093853.5 & 0.239935 & 45.09 & 0.269 & 7.55 & 0.416 \\
SDSS J002053.31+003812.7 & 0.144371 & 44.33 & 0.135 & 7.09 & 0.162 \\
SDSS J002213.00-004832.5 & 0.213941 & 44.54 & 0.195 & 7.14 & $-$0.038 \\
SDSS J002527.35+160226.8 & 0.110662 & 44.26 & 0.131 & 7.03 & 0.084 \\
SDSS J002752.39+002615.7 & 0.205309 & 44.89 & 0.245 & 7.39 & 0.232 \\
SDSS J002830.95-002402.4 & 0.167345 & 44.56 & 0.147 & 7.28 & 0.416 \\
SDSS J002947.82-002258.5 & 0.228873 & 44.81 & 0.228 & 7.34 & 0.210 \\
SDSS J003238.20-010035.2 & 0.091861 & 44.38 & 0.147 & 7.10 & 0.136 \\
SDSS J003542.67+004735.1 & 0.148886 & 44.42 & 0.183 & 7.05 & $-$0.180 \\
SDSS J003644.72+160830.7 & 0.465128 & 45.13 & 0.259 & 7.61 & 0.512 \\
SDSS J003646.45+145936.9 & 0.089210 & 44.05 & 0.125 & 6.84 & $-$0.262 \\
SDSS J003711.00+002127.8 & 0.235101 & 44.96 & 0.235 & 7.48 & 0.400 \\
SDSS J003803.51+145057.3 & 0.138844 & 44.60 & 0.282 & 7.04 & $-$0.686 \\

\bottomrule
\end{tabularx}
\end{table}

\begin{table}[H]\ContinuedFloat
\caption{{\em Cont.}}
\begin{tabularx}{\textwidth}{m{4cm}<{\raggedright}m{1.3cm}<{\centering}m{2.5cm}<{\centering}m{1cm}<{\centering}m{1.3cm}<{\centering}m{1.2cm}<{\centering}}
\toprule
\textbf{Object}	& \textbf{\boldmath$z$} & \textbf{\boldmath$\log(L_\text{bol}\text{[erg/s]})$} & \textbf{\boldmath$l_\text{E}$} & \textbf{\boldmath$M_\text{BH}/M_\odot$} & \textbf{\boldmath$a$}\\
\midrule

SDSS J003846.90+150708.1 & 0.438353 & 45.19 & 0.295 & 7.61 & 0.446 \\
SDSS J004241.90+150926.1 & 0.101278 & 44.36 & 0.151 & 7.07 & 0.056 \\
SDSS J004742.58-004249.7 & 0.147298 & 44.63 & 0.229 & 7.16 & $-$0.142 \\
SDSS J004809.94+151454.5 & 0.114583 & 44.24 & 0.127 & 7.03 & 0.110 \\
SDSS J010009.31+010115.1 & 0.312451 & 44.94 & 0.193 & 7.54 & 0.580 \\
SDSS J010044.84+144535.9 & 0.291464 & 44.90 & 0.295 & 7.32 & $-$0.038 \\
SDSS J010407.00+011412.9 & 0.311880 & 45.26 & 0.275 & 7.71 & 0.590 \\
SDSS J010546.50+000704.9 & 0.263738 & 44.76 & 0.188 & 7.38 & 0.384 \\
SDSS J010950.84+152730.8 & 0.179014 & 44.60 & 0.277 & 7.05 & $-$0.686 \\
SDSS J011009.01-100843.5 & 0.058116 & 44.00 & 0.118 & 6.82 & $-$0.262 \\
SDSS J012801.99+135551.0 & 0.409060 & 44.82 & 0.242 & 7.33 & 0.136 \\
SDSS J013521.68-004402.1 & 0.098500 & 44.55 & 0.184 & 7.17 & 0.084 \\
SDSS J013556.98+003056.9 & 0.137892 & 44.42 & 0.215 & 6.98 & $-$0.564 \\
SDSS J014019.06-092110.4 & 0.135279 & 44.73 & 0.285 & 7.17 & $-$0.354 \\
SDSS J014105.88-100948.1 & 0.126420 & 44.64 & 0.262 & 7.11 & $-$0.402 \\
SDSS J014153.62+125726.9 & 0.198901 & 44.73 & 0.162 & 7.41 & 0.512 \\
SDSS J014248.31-100840.1 & 0.090390 & 44.51 & 0.154 & 7.21 & 0.274 \\
SDSS J014248.86+142125.9 & 0.133879 & 44.64 & 0.286 & 7.07 & $-$0.624 \\
SDSS J014559.44+003524.7 & 0.165335 & 44.45 & 0.140 & 7.19 & 0.314 \\
SDSS J014951.65+002536.4 & 0.252437 & 44.92 & 0.197 & 7.52 & 0.536 \\
SDSS J015046.68+132359.9 & 0.094126 & 44.06 & 0.122 & 6.87 & $-$0.180 \\
SDSS J015219.33+141206.5 & 0.248238 & 45.12 & 0.264 & 7.59 & 0.474 \\
SDSS J015313.07-091418.7 & 0.298273 & 44.88 & 0.287 & 7.31 & $-$0.038 \\
SDSS J020132.56+002353.2 & 0.078002 & 44.10 & 0.089 & 7.04 & 0.384 \\
SDSS J020844.09+140332.9 & 0.361032 & 45.25 & 0.307 & 7.65 & 0.474 \\
SDSS J021529.30-001448.0 & 0.180980 & 44.66 & 0.153 & 7.37 & 0.486 \\
SDSS J022008.96-090410.0 & 0.231675 & 44.97 & 0.229 & 7.50 & 0.446 \\
SDSS J022226.12-085701.3 & 0.166732 & 44.74 & 0.252 & 7.23 & $-$0.070 \\
SDSS J022347.48-083655.7 & 0.260775 & 45.02 & 0.275 & 7.47 & 0.294 \\
SDSS J022821.38-082106.2 & 0.171374 & 44.56 & 0.263 & 7.03 & $-$0.624 \\
SDSS J023038.88-000114.5 & 0.133688 & 44.60 & 0.211 & 7.16 & $-$0.070 \\
SDSS J023315.91-081633.4 & 0.265553 & 44.88 & 0.285 & 7.31 & $-$0.004 \\
SDSS J023414.57+005708.0 & 0.269222 & 44.81 & 0.307 & 7.21 & $-$0.306 \\
SDSS J024225.87-004142.6 & 0.382742 & 44.59 & 0.186 & 7.21 & 0.136 \\
SDSS J024546.10-085842.2 & 0.148299 & 44.87 & 0.210 & 7.44 & 0.400 \\
SDSS J024621.02+001919.0 & 0.182892 & 44.66 & 0.198 & 7.25 & 0.162 \\
SDSS J024727.54-001041.5 & 0.339179 & 45.04 & 0.232 & 7.57 & 0.512 \\
SDSS J024934.59-082742.6 & 0.520684 & 45.23 & 0.284 & 7.67 & 0.536 \\
SDSS J025416.89-084544.0 & 0.302060 & 45.01 & 0.249 & 7.50 & 0.400 \\
SDSS J030639.57+000343.1 & 0.107344 & 44.87 & 0.229 & 7.40 & 0.294 \\
SDSS J031532.27+005503.5 & 0.487399 & 45.13 & 0.265 & 7.60 & 0.486 \\
SDSS J031722.16-065343.0 & 0.156204 & 44.34 & 0.148 & 7.06 & 0.026 \\
SDSS J032801.70+002100.1 & 0.322078 & 44.93 & 0.214 & 7.49 & 0.460 \\
SDSS J033502.22-005637.9 & 0.193281 & 44.69 & 0.242 & 7.20 & $-$0.106 \\
SDSS J034131.94-000933.0 & 0.223370 & 44.66 & 0.252 & 7.15 & $-$0.262 \\
SDSS J035827.45-050535.1 & 0.199733 & 44.53 & 0.145 & 7.26 & 0.384 \\
SDSS J073606.62+375038.8 & 0.262155 & 44.81 & 0.276 & 7.26 & $-$0.106 \\
\bottomrule
\end{tabularx}
\label{table1_1}
\end{table}
\vspace{-8pt}

\begin{table}[H]
\caption{All objects with their estimated spin values. Part 2 of 8.}
\footnotesize
\begin{tabularx}{\textwidth}{m{4cm}<{\raggedright}m{1.3cm}<{\centering}m{2.5cm}<{\centering}m{1cm}<{\centering}m{1.3cm}<{\centering}m{1.2cm}<{\centering}}
\toprule
\textbf{Object}	& \textbf{\boldmath$z$} & \textbf{\boldmath$\log(L_\text{bol}\text{[erg/s]})$} & \textbf{\boldmath$l_\text{E}$} & \textbf{\boldmath$M_\text{BH}/M_\odot$} & \textbf{\boldmath$a$}\\
\midrule
SDSS J073642.77+320915.4 & 0.184817 & 44.91 & 0.224 & 7.45 & 0.384 \\
SDSS J074035.86+402234.6 & 0.177275 & 44.51 & 0.234 & 7.03 & $-$0.506 \\
SDSS J074255.78+234252.4 & 0.336502 & 44.92 & 0.266 & 7.39 & 0.162 \\
SDSS J074548.27+284838.0 & 0.158396 & 44.89 & 0.283 & 7.33 & 0.026 \\
SDSS J074615.57+302400.4 & 0.206192 & 44.50 & 0.169 & 7.16 & 0.136 \\

\bottomrule
\end{tabularx}
\end{table}

\begin{table}[H]\ContinuedFloat
\caption{{\em Cont.}}
\begin{tabularx}{\textwidth}{m{4cm}<{\raggedright}m{1.3cm}<{\centering}m{2.5cm}<{\centering}m{1cm}<{\centering}m{1.3cm}<{\centering}m{1.2cm}<{\centering}}
\toprule
\textbf{Object}	& \textbf{\boldmath$z$} & \textbf{\boldmath$\log(L_\text{bol}\text{[erg/s]})$} & \textbf{\boldmath$l_\text{E}$} & \textbf{\boldmath$M_\text{BH}/M_\odot$} & \textbf{\boldmath$a$}\\
\midrule

SDSS J074906.25+354133.7 & 0.225995 & 44.63 & 0.220 & 7.18 & $-$0.070 \\
SDSS J074940.91+375508.2 & 0.116532 & 44.41 & 0.180 & 7.04 & $-$0.180 \\
SDSS J075141.57+353914.8 & 0.306203 & 45.08 & 0.295 & 7.50 & 0.294 \\
SDSS J075209.09+414235.5 & 0.258357 & 45.10 & 0.295 & 7.52 & 0.314 \\
SDSS J075347.92+240429.7 & 0.320213 & 45.03 & 0.272 & 7.49 & 0.314 \\
SDSS J075433.59+401209.2 & 0.519061 & 44.80 & 0.266 & 7.26 & $-$0.070 \\
SDSS J075525.29+391109.8 & 0.033513 & 44.22 & 0.172 & 6.87 & $-$0.564 \\
SDSS J075606.59+292501.9 & 0.160578 & 44.30 & 0.159 & 6.99 & $-$0.180 \\
SDSS J075613.68+391513.4 & 0.298794 & 45.05 & 0.275 & 7.50 & 0.332 \\
SDSS J075616.70+252410.9 & 0.284962 & 45.03 & 0.209 & 7.60 & 0.598 \\
SDSS J075659.43+240846.0 & 0.376471 & 44.95 & 0.308 & 7.35 & $-$0.004 \\
SDSS J075838.13+414512.4 & 0.093531 & 44.44 & 0.177 & 7.08 & $-$0.070 \\
SDSS J075922.36+332709.0 & 0.137888 & 44.32 & 0.097 & 7.22 & 0.548 \\
SDSS J080135.10+270214.1 & 0.191474 & 44.66 & 0.292 & 7.08 & $-$0.624 \\
SDSS J080203.03+435940.1 & 0.074398 & 44.01 & 0.073 & 7.04 & 0.486 \\
SDSS J080252.91+314226.1 & 0.483675 & 45.14 & 0.265 & 7.61 & 0.500 \\
SDSS J080416.30+292721.6 & 0.443993 & 44.94 & 0.288 & 7.37 & 0.084 \\
SDSS J080439.54+315809.4 & 0.374000 & 45.13 & 0.246 & 7.63 & 0.558 \\
SDSS J080515.99+334548.6 & 0.247069 & 44.60 & 0.229 & 7.13 & $-$0.220 \\
SDSS J080538.22+244214.8 & 0.098837 & 44.24 & 0.186 & 6.86 & $-$0.686 \\
SDSS J080710.87+245105.9 & 0.328634 & 45.08 & 0.251 & 7.57 & 0.474 \\
SDSS J080742.46+375332.1 & 0.229941 & 45.01 & 0.272 & 7.47 & 0.294 \\
SDSS J080807.15+251545.6 & 0.233955 & 44.70 & 0.200 & 7.29 & 0.232 \\
SDSS J081231.43+441620.8 & 0.296883 & 45.33 & 0.299 & 7.74 & 0.590 \\
SDSS J081252.45+402348.8 & 0.188447 & 44.82 & 0.170 & 7.48 & 0.568 \\
SDSS J081321.36+393109.0 & 0.204989 & 44.77 & 0.276 & 7.22 & $-$0.180 \\
SDSS J081345.89+381049.7 & 0.079960 & 44.21 & 0.090 & 7.15 & 0.512 \\
SDSS J081422.67+293419.0 & 0.224970 & 44.71 & 0.189 & 7.32 & 0.314 \\
SDSS J081427.49+031031.3 & 0.436499 & 45.15 & 0.297 & 7.57 & 0.384 \\
SDSS J081427.69+433705.1 & 0.224183 & 45.12 & 0.257 & 7.60 & 0.500 \\
SDSS J081503.09+293649.5 & 0.264458 & 44.96 & 0.191 & 7.57 & 0.608 \\
SDSS J081516.87+460430.8 & 0.041184 & 44.29 & 0.196 & 6.89 & $-$0.686 \\
SDSS J081823.29+304637.9 & 0.170442 & 44.73 & 0.306 & 7.13 & $-$0.506 \\
SDSS J081835.73+285022.4 & 0.077240 & 44.34 & 0.178 & 6.98 & $-$0.306 \\
SDSS J081836.56+364334.9 & 0.106051 & 44.41 & 0.119 & 7.22 & 0.446 \\
SDSS J082050.48+472047.5 & 0.129067 & 44.38 & 0.119 & 7.19 & 0.416 \\
SDSS J082405.19+445246.0 & 0.219632 & 45.14 & 0.280 & 7.58 & 0.430 \\
SDSS J082432.99+514123.3 & 0.111106 & 44.06 & 0.090 & 7.00 & 0.294 \\
SDSS J082447.43+302554.2 & 0.365092 & 45.18 & 0.259 & 7.66 & 0.558 \\
SDSS J082921.18+375227.5 & 0.275902 & 44.88 & 0.198 & 7.47 & 0.486 \\
SDSS J083013.45+520541.7 & 0.302577 & 45.03 & 0.227 & 7.56 & 0.524 \\
SDSS J083052.73+481959.4 & 0.223419 & 44.69 & 0.225 & 7.23 & 0.026 \\
SDSS J083105.42+484231.6 & 0.169707 & 44.77 & 0.177 & 7.41 & 0.474 \\
SDSS J083202.15+461425.7 & 0.045906 & 44.12 & 0.110 & 6.97 & 0.110 \\
SDSS J083225.22+304608.3 & 0.089738 & 44.16 & 0.118 & 6.98 & 0.084 \\
SDSS J083237.43+365613.9 & 0.265764 & 45.21 & 0.262 & 7.68 & 0.580 \\
SDSS J083352.82+394333.9 & 0.244864 & 44.99 & 0.251 & 7.48 & 0.368 \\
SDSS J083454.08+542644.5 & 0.101118 & 44.30 & 0.107 & 7.16 & 0.430 \\
SDSS J083651.68+465333.9 & 0.248637 & 44.83 & 0.305 & 7.24 & $-$0.262 \\
SDSS J083810.01+350642.0 & 0.263217 & 44.84 & 0.228 & 7.37 & 0.254 \\
SDSS J084000.14+443737.9 & 0.125391 & 44.70 & 0.187 & 7.32 & 0.314 \\
SDSS J084314.95+384250.4 & 0.121095 & 44.46 & 0.124 & 7.26 & 0.474 \\
SDSS J084818.23+045643.2 & 0.437673 & 45.05 & 0.247 & 7.55 & 0.460 \\
SDSS J084855.35+422247.3 & 0.202080 & 44.71 & 0.178 & 7.35 & 0.384 \\
SDSS J085026.98+324651.8 & 0.219877 & 44.60 & 0.294 & 7.02 & $-$0.824 \\
SDSS J085039.70+333843.5 & 0.174862 & 44.45 & 0.169 & 7.11 & 0.026 \\
SDSS J085256.69+391734.4 & 0.250059 & 44.90 & 0.271 & 7.36 & 0.110 \\
SDSS J085315.22+340432.6 & 0.189949 & 44.64 & 0.237 & 7.16 & $-$0.180 \\
\bottomrule
\end{tabularx}
\label{table1_2}
\end{table}

\begin{table}[H]
\caption{All objects with their estimated spin values. Part 3 of 8.}
\footnotesize
\begin{tabularx}{\textwidth}{m{4cm}<{\raggedright}m{1.3cm}<{\centering}m{2.5cm}<{\centering}m{1cm}<{\centering}m{1.3cm}<{\centering}m{1.2cm}<{\centering}}
\toprule
\textbf{Object}	& \textbf{\boldmath$z$} & \textbf{\boldmath$\log(L_\text{bol}\text{[erg/s]})$} & \textbf{\boldmath$l_\text{E}$} & \textbf{\boldmath$M_\text{BH}/M_\odot$} & \textbf{\boldmath$a$}\\
\midrule
SDSS J085457.23+544820.5 & 0.255907 & 45.09 & 0.268 & 7.55 & 0.416 \\
SDSS J085613.17+363144.8 & 0.171102 & 44.58 & 0.207 & 7.15 & $-$0.070 \\
SDSS J085655.29+442653.3 & 0.179081 & 44.66 & 0.144 & 7.39 & 0.558 \\
SDSS J085738.56+452513.9 & 0.242305 & 44.74 & 0.176 & 7.39 & 0.446 \\
SDSS J085858.79+533917.8 & 0.242489 & 44.94 & 0.232 & 7.46 & 0.384 \\
SDSS J090005.80+005835.5 & 0.250564 & 44.76 & 0.233 & 7.28 & 0.084 \\
SDSS J090015.28+510800.1 & 0.126035 & 44.72 & 0.202 & 7.30 & 0.232 \\
SDSS J090102.33+424957.4 & 0.213868 & 44.71 & 0.262 & 7.18 & $-$0.220 \\
SDSS J090117.82+043656.1 & 0.319724 & 45.17 & 0.306 & 7.57 & 0.368 \\
SDSS J090210.16+444625.5 & 0.266220 & 45.01 & 0.247 & 7.51 & 0.400 \\
SDSS J090354.72+350959.8 & 0.241401 & 44.79 & 0.262 & 7.26 & $-$0.038 \\
SDSS J090523.95+413107.3 & 0.494868 & 44.90 & 0.207 & 7.47 & 0.460 \\
SDSS J090611.61+510928.8 & 0.097854 & 44.26 & 0.126 & 7.05 & 0.136 \\
SDSS J090720.90+053833.2 & 0.383351 & 45.00 & 0.278 & 7.45 & 0.232 \\
SDSS J090734.53+491911.7 & 0.141860 & 44.52 & 0.122 & 7.32 & 0.558 \\
SDSS J090741.40+500814.1 & 0.208675 & 44.95 & 0.255 & 7.43 & 0.274 \\
SDSS J090927.69+393229.9 & 0.152103 & 44.21 & 0.177 & 6.85 & $-$0.686 \\
SDSS J091034.21+533725.9 & 0.187304 & 44.61 & 0.182 & 7.24 & 0.210 \\
SDSS J091113.39+400111.1 & 0.200499 & 44.84 & 0.291 & 7.27 & $-$0.142 \\
SDSS J091245.77+450046.5 & 0.318832 & 44.91 & 0.199 & 7.50 & 0.512 \\
SDSS J091313.72+365817.2 & 0.107321 & 44.54 & 0.185 & 7.16 & 0.056 \\
SDSS J091400.03+462937.3 & 0.136733 & 44.24 & 0.189 & 6.85 & $-$0.754 \\
SDSS J091508.55+530310.2 & 0.248602 & 45.12 & 0.234 & 7.64 & 0.590 \\
SDSS J091512.23+013412.1 & 0.455930 & 44.72 & 0.199 & 7.31 & 0.254 \\
SDSS J091513.89+571233.2 & 0.195095 & 44.35 & 0.198 & 6.94 & $-$0.564 \\
SDSS J091953.24+595128.8 & 0.215659 & 44.77 & 0.189 & 7.38 & 0.400 \\
SDSS J092019.52+463608.9 & 0.155863 & 44.51 & 0.253 & 7.00 & $-$0.686 \\
SDSS J092050.41+422408.0 & 0.174452 & 44.55 & 0.254 & 7.04 & $-$0.564 \\
SDSS J092057.50+510700.3 & 0.200458 & 44.80 & 0.177 & 7.44 & 0.512 \\
SDSS J092351.17+032231.6 & 0.179751 & 44.85 & 0.292 & 7.27 & $-$0.142 \\
SDSS J092506.70+390708.0 & 0.247477 & 44.97 & 0.283 & 7.41 & 0.162 \\
SDSS J092704.38+563351.4 & 0.219553 & 44.69 & 0.141 & 7.43 & 0.598 \\
SDSS J092810.50+411129.1 & 0.152623 & 44.69 & 0.145 & 7.42 & 0.580 \\
SDSS J093147.74+433119.2 & 0.132046 & 44.39 & 0.132 & 7.16 & 0.294 \\
SDSS J093312.47+611936.3 & 0.122705 & 44.04 & 0.086 & 6.99 & 0.332 \\
SDSS J093607.86+051034.4 & 0.207456 & 44.82 & 0.259 & 7.30 & 0.026 \\
SDSS J093919.90+072755.0 & 0.409697 & 45.14 & 0.308 & 7.54 & 0.314 \\
SDSS J094109.86+081404.5 & 0.125429 & 44.22 & 0.148 & 6.94 & $-$0.220 \\
SDSS J094456.07+054642.9 & 0.212865 & 44.90 & 0.290 & 7.33 & $-$0.004 \\
SDSS J094529.36+093610.4 & 0.013277 & 43.12 & 0.055 & 6.27 & $-$0.898 \\
SDSS J094616.90+025459.4 & 0.117764 & 44.60 & 0.192 & 7.21 & 0.110 \\
SDSS J094621.26+471131.3 & 0.230494 & 45.13 & 0.289 & 7.56 & 0.384 \\
SDSS J094842.67+502931.4 & 0.056464 & 44.47 & 0.162 & 7.15 & 0.136 \\
SDSS J094903.55+474653.9 & 0.214958 & 44.75 & 0.252 & 7.24 & $-$0.070 \\
SDSS J095017.60+070317.9 & 0.108707 & 44.30 & 0.175 & 6.95 & $-$0.402 \\
SDSS J095221.96+632438.9 & 0.119716 & 44.33 & 0.115 & 7.16 & 0.384 \\
SDSS J095310.69+032725.5 & 0.184362 & 44.53 & 0.125 & 7.32 & 0.548 \\
SDSS J095553.14+633742.7 & 0.356415 & 45.03 & 0.238 & 7.54 & 0.474 \\
SDSS J095730.15+413301.6 & 0.318701 & 45.02 & 0.255 & 7.50 & 0.384 \\
SDSS J095931.67+504449.0 & 0.143226 & 45.08 & 0.300 & 7.49 & 0.254 \\
SDSS J100201.77+620816.3 & 0.133791 & 44.56 & 0.226 & 7.10 & $-$0.262 \\
SDSS J100706.25+084228.4 & 0.373343 & 45.19 & 0.300 & 7.60 & 0.416 \\
SDSS J100723.15+014546.8 & 0.355111 & 44.73 & 0.266 & 7.19 & $-$0.220 \\
SDSS J100954.65+481514.6 & 0.178805 & 44.65 & 0.169 & 7.31 & 0.368 \\
SDSS J101341.90-000925.7 & 0.277105 & 44.98 & 0.267 & 7.44 & 0.254 \\
SDSS J101437.45+440639.1 & 0.200135 & 45.16 & 0.270 & 7.62 & 0.500 \\

\bottomrule
\end{tabularx}
\end{table}

\begin{table}[H]\ContinuedFloat
\caption{{\em Cont.}}
\begin{tabularx}{\textwidth}{m{4cm}<{\raggedright}m{1.3cm}<{\centering}m{2.5cm}<{\centering}m{1cm}<{\centering}m{1.3cm}<{\centering}m{1.2cm}<{\centering}}
\toprule
\textbf{Object}	& \textbf{\boldmath$z$} & \textbf{\boldmath$\log(L_\text{bol}\text{[erg/s]})$} & \textbf{\boldmath$l_\text{E}$} & \textbf{\boldmath$M_\text{BH}/M_\odot$} & \textbf{\boldmath$a$}\\
\midrule

SDSS J101549.33+424243.0 & 0.498800 & 45.33 & 0.296 & 7.75 & 0.598 \\
SDSS J101645.11+421025.5 & 0.055322 & 44.58 & 0.280 & 7.02 & $-$0.754 \\
SDSS J101852.45+495800.4 & 0.154811 & 44.63 & 0.208 & 7.20 & 0.026 \\
SDSS J101936.27+002029.7 & 0.147878 & 44.61 & 0.250 & 7.10 & $-$0.402 \\
SDSS J102000.45+623944.6 & 0.136020 & 44.44 & 0.211 & 7.01 & $-$0.452 \\
SDSS J102049.68+060446.9 & 0.110954 & 44.41 & 0.154 & 7.11 & 0.110 \\
SDSS J102148.89+030732.2 & 0.061838 & 44.25 & 0.143 & 6.98 & $-$0.070 \\
\bottomrule
\end{tabularx}
\label{table1_3}
\end{table}

\vspace{-8pt}

\begin{table}[H]
\caption{All objects with their estimated spin values. Part 4 of 8.}
\footnotesize
\begin{tabularx}{\textwidth}{m{4cm}<{\raggedright}m{1.3cm}<{\centering}m{2.5cm}<{\centering}m{1cm}<{\centering}m{1.3cm}<{\centering}m{1.2cm}<{\centering}}
\toprule
\textbf{Object}	& \textbf{\boldmath$z$} & \textbf{\boldmath$\log(L_\text{bol}\text{[erg/s]})$} & \textbf{\boldmath$l_\text{E}$} & \textbf{\boldmath$M_\text{BH}/M_\odot$} & \textbf{\boldmath$a$}\\
\midrule
SDSS J102307.02+454500.1 & 0.256396 & 44.73 & 0.166 & 7.40 & 0.486 \\
SDSS J102402.59+062943.9 & 0.044014 & 43.94 & 0.117 & 6.76 & $-$0.402 \\
SDSS J102448.57+003538.0 & 0.095475 & 44.33 & 0.154 & 7.03 & $-$0.038 \\
SDSS J102531.28+514034.8 & 0.044884 & 44.47 & 0.242 & 6.98 & $-$0.686 \\
SDSS J102712.37+050320.9 & 0.338470 & 44.98 & 0.280 & 7.42 & 0.186 \\
SDSS J102748.92+651337.4 & 0.444910 & 44.98 & 0.279 & 7.42 & 0.210 \\
SDSS J102754.27+063107.4 & 0.159051 & 44.50 & 0.217 & 7.05 & $-$0.354 \\
SDSS J102905.51+052555.3 & 0.122770 & 44.41 & 0.132 & 7.18 & 0.332 \\
SDSS J102955.38+471346.6 & 0.175436 & 44.78 & 0.218 & 7.33 & 0.232 \\
SDSS J103103.52+462616.8 & 0.093244 & 44.48 & 0.171 & 7.14 & 0.056 \\
SDSS J104153.59+031500.6 & 0.093442 & 44.27 & 0.099 & 7.16 & 0.474 \\
SDSS J104241.08+520012.8 & 0.136249 & 44.63 & 0.242 & 7.14 & $-$0.262 \\
SDSS J104331.50-010732.8 & 0.361906 & 45.15 & 0.274 & 7.60 & 0.474 \\
SDSS J104413.32-000324.7 & 0.279814 & 45.13 & 0.291 & 7.56 & 0.384 \\
SDSS J104840.49+541302.3 & 0.104141 & 44.28 & 0.194 & 6.88 & $-$0.686 \\
SDSS J105237.41+503042.1 & 0.245730 & 45.08 & 0.259 & 7.56 & 0.446 \\
SDSS J110501.99-004454.7 & 0.332404 & 45.08 & 0.238 & 7.59 & 0.536 \\
SDSS J110522.54+510727.1 & 0.307074 & 45.05 & 0.288 & 7.48 & 0.274 \\
SDSS J110608.03+582609.1 & 0.118018 & 44.45 & 0.217 & 7.00 & $-$0.506 \\
SDSS J110711.32+080538.4 & 0.281045 & 45.08 & 0.233 & 7.60 & 0.558 \\
SDSS J110735.68+060758.6 & 0.380035 & 45.19 & 0.290 & 7.62 & 0.460 \\
SDSS J111012.07+011327.8 & 0.094936 & 45.10 & 0.302 & 7.51 & 0.294 \\
SDSS J111233.81+034007.0 & 0.321768 & 44.86 & 0.223 & 7.40 & 0.314 \\
SDSS J111407.35-000031.1 & 0.072649 & 44.14 & 0.136 & 6.90 & $-$0.220 \\
SDSS J111528.48-003454.7 & 0.232516 & 44.69 & 0.199 & 7.28 & 0.210 \\
SDSS J111928.85+042655.5 & 0.153477 & 44.56 & 0.167 & 7.23 & 0.254 \\
SDSS J111949.75+020256.5 & 0.207822 & 44.60 & 0.221 & 7.15 & $-$0.142 \\
SDSS J112016.16+491428.8 & 0.149581 & 44.40 & 0.208 & 6.97 & $-$0.506 \\
SDSS J112108.58+535121.0 & 0.102907 & 45.27 & 0.298 & 7.69 & 0.524 \\
SDSS J112114.22+032546.7 & 0.152033 & 45.07 & 0.267 & 7.53 & 0.400 \\
SDSS J112209.40+011719.3 & 0.058292 & 44.08 & 0.164 & 6.76 & $-$0.898 \\
SDSS J112328.12+052823.2 & 0.101336 & 44.42 & 0.145 & 7.15 & 0.232 \\
SDSS J112606.43+002349.9 & 0.169536 & 44.85 & 0.169 & 7.51 & 0.608 \\
SDSS J112747.17+632542.7 & 0.341665 & 44.54 & 0.209 & 7.11 & $-$0.180 \\
SDSS J112805.72-005850.9 & 0.284739 & 45.04 & 0.226 & 7.58 & 0.536 \\
SDSS J112836.17+024550.6 & 0.238576 & 44.65 & 0.168 & 7.31 & 0.368 \\
SDSS J113001.88+494434.7 & 0.244338 & 45.29 & 0.301 & 7.70 & 0.536 \\
SDSS J113003.11+655629.1 & 0.132687 & 44.62 & 0.276 & 7.07 & $-$0.564 \\
SDSS J113102.27-010122.0 & 0.242137 & 44.52 & 0.256 & 7.00 & $-$0.686 \\
SDSS J113110.64+043856.0 & 0.144594 & 44.66 & 0.184 & 7.29 & 0.274 \\
SDSS J113111.94+100231.3 & 0.074401 & 44.05 & 0.112 & 6.89 & $-$0.070 \\
SDSS J113151.04+100915.5 & 0.119481 & 44.31 & 0.213 & 6.87 & $-$0.824 \\
SDSS J113223.43+641958.4 & 0.209867 & 45.17 & 0.261 & 7.64 & 0.548 \\

\bottomrule
\end{tabularx}
\end{table}

\begin{table}[H]\ContinuedFloat
\caption{{\em Cont.}}
\begin{tabularx}{\textwidth}{m{4cm}<{\raggedright}m{1.3cm}<{\centering}m{2.5cm}<{\centering}m{1cm}<{\centering}m{1.3cm}<{\centering}m{1.2cm}<{\centering}}
\toprule
\textbf{Object}	& \textbf{\boldmath$z$} & \textbf{\boldmath$\log(L_\text{bol}\text{[erg/s]})$} & \textbf{\boldmath$l_\text{E}$} & \textbf{\boldmath$M_\text{BH}/M_\odot$} & \textbf{\boldmath$a$}\\
\midrule

SDSS J113229.54+092042.0 & 0.305157 & 45.04 & 0.214 & 7.60 & 0.590 \\
SDSS J113320.91+043255.1 & 0.248058 & 45.15 & 0.241 & 7.66 & 0.598 \\
SDSS J113842.84-031403.3 & 0.212339 & 44.92 & 0.210 & 7.49 & 0.474 \\
SDSS J113900.50+591347.2 & 0.115052 & 44.68 & 0.187 & 7.30 & 0.274 \\
SDSS J114203.66+054850.4 & 0.274033 & 44.86 & 0.198 & 7.45 & 0.460 \\
SDSS J114208.48+531526.0 & 0.067965 & 44.20 & 0.097 & 7.10 & 0.416 \\
SDSS J114341.97-014434.4 & 0.105223 & 44.93 & 0.275 & 7.38 & 0.136 \\
SDSS J114514.00+494523.4 & 0.192336 & 44.56 & 0.261 & 7.03 & $-$0.624 \\
SDSS J114632.86+030506.9 & 0.191294 & 44.45 & 0.179 & 7.09 & $-$0.070 \\
SDSS J114928.27-000442.5 & 0.175967 & 44.67 & 0.149 & 7.39 & 0.536 \\
SDSS J114958.08+575107.7 & 0.100556 & 44.44 & 0.244 & 6.94 & $-$0.824 \\
SDSS J115050.20+004505.7 & 0.139474 & 44.52 & 0.128 & 7.30 & 0.512 \\
SDSS J115215.83+042456.2 & 0.132674 & 44.30 & 0.136 & 7.06 & 0.110 \\
SDSS J115655.88+084850.2 & 0.496092 & 45.26 & 0.275 & 7.71 & 0.590 \\
SDSS J115713.04+535312.8 & 0.272882 & 44.84 & 0.254 & 7.32 & 0.084 \\
SDSS J115715.19+093456.7 & 0.271835 & 44.79 & 0.160 & 7.48 & 0.590 \\
SDSS J115723.17+045201.0 & 0.176083 & 44.43 & 0.211 & 7.00 & $-$0.452 \\
SDSS J115741.75+041250.6 & 0.094812 & 44.29 & 0.148 & 7.01 & $-$0.070 \\
SDSS J115755.47+001704.0 & 0.260770 & 44.93 & 0.227 & 7.46 & 0.400 \\
SDSS J115852.57+563152.5 & 0.234345 & 44.60 & 0.270 & 7.06 & $-$0.624 \\
\bottomrule
\end{tabularx}
\label{table1_4}
\end{table}
\vspace{-8pt}

\begin{table}[H]
\caption{All objects with their estimated spin values. Part 5 of 8.}
\footnotesize
\begin{tabularx}{\textwidth}{m{4cm}<{\raggedright}m{1.3cm}<{\centering}m{2.5cm}<{\centering}m{1cm}<{\centering}m{1.3cm}<{\centering}m{1.2cm}<{\centering}}
\toprule
\textbf{Object}	& \textbf{\boldmath$z$} & \textbf{\boldmath$\log(L_\text{bol}\text{[erg/s]})$} & \textbf{\boldmath$l_\text{E}$} & \textbf{\boldmath$M_\text{BH}/M_\odot$} & \textbf{\boldmath$a$}\\
\midrule
SDSS J115905.80+024802.6 & 0.168560 & 44.65 & 0.216 & 7.21 & $-$0.004 \\
SDSS J120014.08-004638.7 & 0.179389 & 44.93 & 0.236 & 7.45 & 0.350 \\
SDSS J120322.36+621505.7 & 0.270417 & 44.82 & 0.303 & 7.23 & $-$0.262 \\
SDSS J120517.71+520109.1 & 0.189541 & 44.72 & 0.210 & 7.29 & 0.186 \\
SDSS J120628.97+503001.5 & 0.171971 & 44.58 & 0.272 & 7.03 & $-$0.686 \\
SDSS J121117.80-000212.4 & 0.181317 & 44.72 & 0.177 & 7.36 & 0.416 \\
SDSS J121157.48+055801.1 & 0.067820 & 44.18 & 0.136 & 6.94 & $-$0.142 \\
SDSS J121255.27+512221.1 & 0.282823 & 44.85 & 0.241 & 7.36 & 0.210 \\
SDSS J121333.20-013220.7 & 0.197743 & 44.63 & 0.204 & 7.21 & 0.056 \\
SDSS J121343.76-010002.5 & 0.328065 & 44.90 & 0.239 & 7.41 & 0.294 \\
SDSS J121407.35+655228.6 & 0.235552 & 45.10 & 0.238 & 7.61 & 0.558 \\
SDSS J121513.83+023334.4 & 0.224435 & 44.72 & 0.157 & 7.41 & 0.536 \\
SDSS J121544.73+592639.1 & 0.095815 & 44.21 & 0.122 & 7.01 & 0.110 \\
SDSS J121948.93+054531.7 & 0.113866 & 44.48 & 0.166 & 7.15 & 0.110 \\
SDSS J122342.82+581446.2 & 0.014527 & 42.96 & 0.042 & 6.22 & $-$0.686 \\
SDSS J122450.55+100545.4 & 0.167900 & 44.91 & 0.244 & 7.41 & 0.274 \\
SDSS J122506.20-030100.4 & 0.239966 & 44.70 & 0.297 & 7.12 & $-$0.564 \\
SDSS J122624.42+014020.7 & 0.219884 & 44.83 & 0.176 & 7.47 & 0.548 \\
SDSS J122801.33+623948.1 & 0.271290 & 44.99 & 0.251 & 7.48 & 0.368 \\
SDSS J122908.95+561109.1 & 0.265475 & 44.78 & 0.180 & 7.41 & 0.460 \\
SDSS J122950.61+024652.7 & 0.336083 & 45.15 & 0.284 & 7.59 & 0.430 \\
SDSS J123003.50+611904.7 & 0.148374 & 44.73 & 0.276 & 7.18 & $-$0.306 \\
SDSS J123012.17+544719.8 & 0.276800 & 44.82 & 0.201 & 7.41 & 0.400 \\
SDSS J123132.52+574624.8 & 0.259910 & 44.85 & 0.218 & 7.40 & 0.332 \\
SDSS J123339.58+052034.7 & 0.215649 & 44.64 & 0.226 & 7.18 & $-$0.106 \\
SDSS J123340.07+680022.4 & 0.343372 & 44.96 & 0.244 & 7.46 & 0.350 \\
SDSS J123450.50+040845.4 & 0.121252 & 44.63 & 0.244 & 7.13 & $-$0.262 \\
SDSS J123831.33+643456.5 & 0.101584 & 44.73 & 0.221 & 7.28 & 0.110 \\
SDSS J124110.10+104143.7 & 0.156232 & 44.24 & 0.168 & 6.91 & $-$0.452 \\
SDSS J124129.34+681533.9 & 0.150972 & 44.24 & 0.128 & 7.02 & 0.084 \\
SDSS J124328.04+565237.9 & 0.106616 & 44.29 & 0.165 & 6.96 & $-$0.262 \\

\bottomrule
\end{tabularx}
\end{table}

\begin{table}[H]\ContinuedFloat
\caption{{\em Cont.}}
\begin{tabularx}{\textwidth}{m{4cm}<{\raggedright}m{1.3cm}<{\centering}m{2.5cm}<{\centering}m{1cm}<{\centering}m{1.3cm}<{\centering}m{1.2cm}<{\centering}}
\toprule
\textbf{Object}	& \textbf{\boldmath$z$} & \textbf{\boldmath$\log(L_\text{bol}\text{[erg/s]})$} & \textbf{\boldmath$l_\text{E}$} & \textbf{\boldmath$M_\text{BH}/M_\odot$} & \textbf{\boldmath$a$}\\
\midrule

SDSS J124504.57+650122.7 & 0.206591 & 44.66 & 0.295 & 7.08 & $-$0.624 \\
SDSS J124504.93+504446.2 & 0.129684 & 44.59 & 0.209 & 7.16 & $-$0.070 \\
SDSS J124519.73-005230.5 & 0.221020 & 44.61 & 0.234 & 7.13 & $-$0.220 \\
SDSS J125051.04+060910.0 & 0.182047 & 45.14 & 0.308 & 7.54 & 0.314 \\
SDSS J125156.50+015249.6 & 0.329416 & 45.31 & 0.292 & 7.73 & 0.590 \\
SDSS J125224.22+645901.4 & 0.220729 & 44.66 & 0.245 & 7.16 & $-$0.220 \\
SDSS J125227.32+032353.6 & 0.132687 & 44.81 & 0.233 & 7.33 & 0.186 \\
SDSS J125248.49+015236.3 & 0.287988 & 44.84 & 0.178 & 7.48 & 0.548 \\
SDSS J125357.41+640534.8 & 0.267455 & 44.71 & 0.258 & 7.19 & $-$0.220 \\
SDSS J125635.87+500852.3 & 0.245337 & 44.67 & 0.203 & 7.25 & 0.136 \\
SDSS J130030.67+485042.5 & 0.251725 & 44.90 & 0.190 & 7.51 & 0.548 \\
SDSS J130052.10+564105.9 & 0.071838 & 44.41 & 0.126 & 7.20 & 0.384 \\
SDSS J130421.89+014915.9 & 0.153597 & 44.53 & 0.217 & 7.08 & $-$0.262 \\
SDSS J130547.00+504034.0 & 0.055124 & 43.92 & 0.109 & 6.77 & $-$0.306 \\
SDSS J130717.75+033447.5 & 0.161346 & 44.30 & 0.143 & 7.03 & 0.026 \\
SDSS J131136.37+580801.5 & 0.071088 & 44.16 & 0.087 & 7.11 & 0.486 \\
SDSS J131234.32+655240.1 & 0.217351 & 44.45 & 0.236 & 6.97 & $-$0.686 \\
SDSS J131305.81+012755.9 & 0.029361 & 43.71 & 0.088 & 6.65 & $-$0.354 \\
SDSS J132026.49+051113.5 & 0.098391 & 44.51 & 0.238 & 7.02 & $-$0.506 \\
SDSS J132231.12-001124.6 & 0.172930 & 44.67 & 0.147 & 7.39 & 0.536 \\
SDSS J132428.34+590423.8 & 0.239623 & 44.81 & 0.247 & 7.31 & 0.084 \\
SDSS J132447.09+530257.6 & 0.292005 & 44.90 & 0.234 & 7.42 & 0.314 \\
SDSS J132640.04+650427.4 & 0.400867 & 45.07 & 0.290 & 7.50 & 0.294 \\
SDSS J132704.54-003627.5 & 0.301717 & 44.90 & 0.268 & 7.36 & 0.110 \\
SDSS J132705.88-012415.5 & 0.167808 & 44.81 & 0.272 & 7.26 & $-$0.070 \\
SDSS J132731.98+654848.3 & 0.219769 & 44.72 & 0.187 & 7.34 & 0.350 \\
SDSS J133059.07+602128.4 & 0.291747 & 45.04 & 0.223 & 7.58 & 0.548 \\
SDSS J133138.03+013151.7 & 0.080473 & 44.42 & 0.159 & 7.11 & 0.084 \\
SDSS J133248.59+442452.7 & 0.077438 & 44.13 & 0.101 & 7.02 & 0.254 \\
SDSS J133315.25+560859.8 & 0.343105 & 45.21 & 0.263 & 7.68 & 0.580 \\
SDSS J133328.96+613513.3 & 0.151516 & 44.67 & 0.141 & 7.41 & 0.580 \\
SDSS J133627.97+442917.7 & 0.137809 & 44.56 & 0.243 & 7.06 & $-$0.452 \\
\bottomrule
\end{tabularx}
\label{table1_5}
\end{table}
\vspace{-8pt}

\begin{table}[H]
\caption{All objects with their estimated spin values. Part 6 of 8.}
\footnotesize
\begin{tabularx}{\textwidth}{m{4cm}<{\raggedright}m{1.3cm}<{\centering}m{2.5cm}<{\centering}m{1cm}<{\centering}m{1.3cm}<{\centering}m{1.2cm}<{\centering}}
\toprule
\textbf{Object}	& \textbf{\boldmath$z$} & \textbf{\boldmath$\log(L_\text{bol}\text{[erg/s]})$} & \textbf{\boldmath$l_\text{E}$} & \textbf{\boldmath$M_\text{BH}/M_\odot$} & \textbf{\boldmath$a$}\\
\midrule
SDSS J133729.04+563907.6 & 0.143501 & 44.61 & 0.213 & 7.17 & $-$0.070 \\
SDSS J134313.40+654110.4 & 0.240726 & 44.90 & 0.231 & 7.43 & 0.332 \\
SDSS J134351.06+000434.7 & 0.073693 & 44.36 & 0.104 & 7.23 & 0.536 \\
SDSS J134452.91+000520.2 & 0.087099 & 44.42 & 0.124 & 7.22 & 0.416 \\
SDSS J134524.69-025939.8 & 0.085402 & 44.59 & 0.156 & 7.29 & 0.384 \\
SDSS J134730.70+603742.8 & 0.143551 & 44.56 & 0.146 & 7.28 & 0.416 \\
SDSS J135343.63-011801.3 & 0.145193 & 44.53 & 0.168 & 7.19 & 0.186 \\
SDSS J135350.63+571725.8 & 0.234458 & 44.81 & 0.274 & 7.26 & $-$0.106 \\
SDSS J135622.94+574150.9 & 0.309698 & 44.96 & 0.292 & 7.38 & 0.110 \\
SDSS J135643.69+664128.4 & 0.172599 & 44.59 & 0.257 & 7.07 & $-$0.506 \\
SDSS J135756.53+655902.9 & 0.197031 & 44.83 & 0.284 & 7.27 & $-$0.106 \\
SDSS J135842.27+024925.1 & 0.148016 & 44.56 & 0.170 & 7.22 & 0.232 \\
SDSS J135848.54+430435.6 & 0.252436 & 44.81 & 0.242 & 7.32 & 0.110 \\
SDSS J135944.07+045649.5 & 0.085623 & 44.25 & 0.169 & 6.91 & $-$0.452 \\
SDSS J140046.05+531920.2 & 0.388442 & 45.17 & 0.285 & 7.61 & 0.460 \\
SDSS J140219.69+521059.4 & 0.279175 & 44.88 & 0.271 & 7.34 & 0.056 \\
SDSS J140322.10+022232.9 & 0.250159 & 44.66 & 0.308 & 7.06 & $-$0.754 \\
SDSS J140527.68+505546.5 & 0.106561 & 44.53 & 0.153 & 7.24 & 0.314 \\
SDSS J140926.77+473127.3 & 0.143403 & 44.41 & 0.220 & 6.96 & $-$0.624 \\

\bottomrule
\end{tabularx}
\end{table}

\begin{table}[H]\ContinuedFloat
\caption{{\em Cont.}}
\begin{tabularx}{\textwidth}{m{4cm}<{\raggedright}m{1.3cm}<{\centering}m{2.5cm}<{\centering}m{1cm}<{\centering}m{1.3cm}<{\centering}m{1.2cm}<{\centering}}
\toprule
\textbf{Object}	& \textbf{\boldmath$z$} & \textbf{\boldmath$\log(L_\text{bol}\text{[erg/s]})$} & \textbf{\boldmath$l_\text{E}$} & \textbf{\boldmath$M_\text{BH}/M_\odot$} & \textbf{\boldmath$a$}\\
\midrule

SDSS J141108.51+424428.9 & 0.173315 & 44.58 & 0.170 & 7.24 & 0.254 \\
SDSS J141419.84+533815.3 & 0.164455 & 44.98 & 0.269 & 7.44 & 0.254 \\
SDSS J141424.90+465348.5 & 0.149731 & 44.79 & 0.220 & 7.34 & 0.232 \\
SDSS J141820.32-005953.8 & 0.253638 & 44.94 & 0.269 & 7.40 & 0.186 \\
SDSS J141838.27+620718.5 & 0.138792 & 44.59 & 0.134 & 7.35 & 0.548 \\
SDSS J142103.52+515819.4 & 0.263543 & 44.59 & 0.130 & 7.37 & 0.568 \\
SDSS J142214.89+431357.4 & 0.323267 & 45.09 & 0.263 & 7.56 & 0.446 \\
SDSS J142509.12+011911.4 & 0.199814 & 44.50 & 0.192 & 7.11 & $-$0.106 \\
SDSS J142542.55+652716.9 & 0.241476 & 44.77 & 0.259 & 7.25 & $-$0.070 \\
SDSS J142830.16+555931.3 & 0.351431 & 45.14 & 0.252 & 7.63 & 0.548 \\
SDSS J143030.21-001115.0 & 0.103284 & 44.12 & 0.100 & 7.01 & 0.254 \\
SDSS J143223.67+400533.8 & 0.140651 & 44.47 & 0.158 & 7.16 & 0.162 \\
SDSS J143249.68+451338.2 & 0.306872 & 45.07 & 0.308 & 7.47 & 0.210 \\
SDSS J143407.20+452732.2 & 0.254952 & 44.67 & 0.269 & 7.13 & $-$0.402 \\
SDSS J143601.55+044807.7 & 0.194125 & 44.90 & 0.272 & 7.36 & 0.110 \\
SDSS J143704.11+000705.1 & 0.140363 & 44.80 & 0.229 & 7.33 & 0.186 \\
SDSS J143715.12+545243.8 & 0.252154 & 45.00 & 0.270 & 7.46 & 0.274 \\
SDSS J143952.91+392358.9 & 0.111999 & 44.36 & 0.198 & 6.95 & $-$0.506 \\
SDSS J144013.89-015708.3 & 0.463581 & 44.83 & 0.197 & 7.43 & 0.430 \\
SDSS J144205.04+545904.7 & 0.104628 & 44.17 & 0.112 & 7.01 & 0.162 \\
SDSS J144237.72+542851.4 & 0.155188 & 44.67 & 0.259 & 7.15 & $-$0.306 \\
SDSS J144249.70+611137.8 & 0.047858 & 43.42 & 0.056 & 6.56 & $-$0.106 \\
SDSS J144328.40+542933.1 & 0.226942 & 44.79 & 0.216 & 7.35 & 0.254 \\
SDSS J144507.31+593649.8 & 0.128030 & 43.88 & 0.104 & 6.75 & $-$0.306 \\
SDSS J144705.46+003653.2 & 0.095493 & 43.84 & 0.088 & 6.78 & $-$0.070 \\
SDSS J144920.25+553429.4 & 0.468295 & 45.17 & 0.278 & 7.62 & 0.474 \\
SDSS J144945.69+422243.2 & 0.262760 & 45.08 & 0.245 & 7.58 & 0.512 \\
SDSS J145123.01-000625.8 & 0.138613 & 44.53 & 0.131 & 7.30 & 0.500 \\
SDSS J145201.55+025335.1 & 0.433702 & 45.12 & 0.286 & 7.55 & 0.384 \\
SDSS J145235.27+495142.1 & 0.144921 & 44.53 & 0.120 & 7.34 & 0.580 \\
SDSS J145624.00+421800.2 & 0.189834 & 44.88 & 0.187 & 7.50 & 0.548 \\
SDSS J145643.88+503756.4 & 0.131486 & 44.56 & 0.244 & 7.06 & $-$0.452 \\
SDSS J145801.49+544056.1 & 0.144858 & 44.40 & 0.129 & 7.18 & 0.332 \\
SDSS J145921.17+521749.6 & 0.167019 & 44.72 & 0.153 & 7.43 & 0.558 \\
SDSS J150034.45+465234.1 & 0.298022 & 45.04 & 0.253 & 7.53 & 0.416 \\
SDSS J150238.69+501524.0 & 0.172268 & 44.61 & 0.195 & 7.21 & 0.084 \\
SDSS J150346.94+420323.1 & 0.168316 & 44.74 & 0.212 & 7.30 & 0.210 \\
SDSS J150816.95+520541.7 & 0.193395 & 44.76 & 0.258 & 7.24 & $-$0.106 \\
SDSS J150832.91+583422.4 & 0.502172 & 45.07 & 0.249 & 7.56 & 0.474 \\
SDSS J151020.05+554722.0 & 0.149693 & 44.53 & 0.259 & 7.01 & $-$0.686 \\
SDSS J151024.93+005844.0 & 0.072262 & 44.41 & 0.153 & 7.11 & 0.110 \\
SDSS J151101.89+520350.0 & 0.211340 & 45.07 & 0.257 & 7.55 & 0.446 \\
SDSS J151131.33+502219.0 & 0.219845 & 45.02 & 0.222 & 7.56 & 0.536 \\
SDSS J151616.18+463515.3 & 0.208271 & 44.54 & 0.242 & 7.05 & $-$0.506 \\
\bottomrule
\end{tabularx}
\label{table1_6}
\end{table}
\vspace{-8pt}

\begin{table}[H]
\caption{All objects with their estimated spin values. Part 7 of 8.}
\footnotesize
\begin{tabularx}{\textwidth}{m{4cm}<{\raggedright}m{1.3cm}<{\centering}m{2.5cm}<{\centering}m{1cm}<{\centering}m{1.3cm}<{\centering}m{1.2cm}<{\centering}}
\toprule
\textbf{Object}	& \textbf{\boldmath$z$} & \textbf{\boldmath$\log(L_\text{bol}\text{[erg/s]})$} & \textbf{\boldmath$l_\text{E}$} & \textbf{\boldmath$M_\text{BH}/M_\odot$} & \textbf{\boldmath$a$}\\
\midrule
SDSS J151617.16+472805.0 & 0.197851 & 44.62 & 0.227 & 7.15 & $-$0.142 \\
SDSS J151956.57+001614.6 & 0.114398 & 44.69 & 0.257 & 7.17 & $-$0.220 \\
SDSS J152209.56+451124.0 & 0.065732 & 44.02 & 0.098 & 6.92 & 0.110 \\
SDSS J152224.45-010838.4 & 0.321221 & 45.21 & 0.257 & 7.69 & 0.598 \\
SDSS J152342.49+033147.9 & 0.221423 & 44.53 & 0.235 & 7.05 & $-$0.452 \\
SDSS J152447.13+520759.1 & 0.160840 & 44.49 & 0.214 & 7.05 & $-$0.354 \\
SDSS J152526.40+400914.3 & 0.355817 & 45.13 & 0.256 & 7.61 & 0.512 \\
SDSS J152621.69+432349.5 & 0.155615 & 44.92 & 0.270 & 7.38 & 0.136 \\

\bottomrule
\end{tabularx}
\end{table}

\begin{table}[H]\ContinuedFloat
\caption{{\em Cont.}}
\begin{tabularx}{\textwidth}{m{4cm}<{\raggedright}m{1.3cm}<{\centering}m{2.5cm}<{\centering}m{1cm}<{\centering}m{1.3cm}<{\centering}m{1.2cm}<{\centering}}
\toprule
\textbf{Object}	& \textbf{\boldmath$z$} & \textbf{\boldmath$\log(L_\text{bol}\text{[erg/s]})$} & \textbf{\boldmath$l_\text{E}$} & \textbf{\boldmath$M_\text{BH}/M_\odot$} & \textbf{\boldmath$a$}\\
\midrule

SDSS J152628.19-003809.4 & 0.123334 & 44.77 & 0.162 & 7.45 & 0.558 \\
SDSS J152840.26+383525.9 & 0.152598 & 44.30 & 0.120 & 7.11 & 0.294 \\
SDSS J152843.94+000740.6 & 0.094440 & 44.25 & 0.157 & 6.94 & $-$0.262 \\
SDSS J152912.14+031815.4 & 0.169956 & 44.35 & 0.095 & 7.26 & 0.608 \\
SDSS J153006.30+010626.0 & 0.239339 & 44.56 & 0.174 & 7.21 & 0.186 \\
SDSS J153252.95+384330.5 & 0.134003 & 44.63 & 0.246 & 7.13 & $-$0.306 \\
SDSS J153458.50+024214.0 & 0.389458 & 45.10 & 0.294 & 7.52 & 0.314 \\
SDSS J153607.72+364806.8 & 0.277578 & 45.12 & 0.266 & 7.59 & 0.474 \\
SDSS J153651.27+541442.6 & 0.366706 & 45.08 & 0.306 & 7.48 & 0.232 \\
SDSS J153705.95+005522.8 & 0.136419 & 45.09 & 0.289 & 7.52 & 0.332 \\
SDSS J153937.82+374340.4 & 0.165048 & 44.54 & 0.155 & 7.24 & 0.314 \\
SDSS J154113.94+492034.5 & 0.308087 & 44.87 & 0.222 & 7.41 & 0.350 \\
SDSS J154623.61+475122.3 & 0.103287 & 44.21 & 0.145 & 6.94 & $-$0.180 \\
SDSS J154656.62+005719.6 & 0.211019 & 44.92 & 0.301 & 7.33 & $-$0.038 \\
SDSS J154814.75+450027.7 & 0.037268 & 43.35 & 0.053 & 6.52 & $-$0.142 \\
SDSS J155427.26+404441.3 & 0.116805 & 44.43 & 0.139 & 7.18 & 0.294 \\
SDSS J155451.13+461917.3 & 0.116881 & 44.44 & 0.213 & 7.00 & $-$0.452 \\
SDSS J155637.99+540308.3 & 0.203331 & 44.86 & 0.240 & 7.37 & 0.232 \\
SDSS J155755.23+331625.8 & 0.277693 & 45.03 & 0.294 & 7.45 & 0.210 \\
SDSS J155851.33+280719.6 & 0.282557 & 45.02 & 0.262 & 7.49 & 0.350 \\
SDSS J155904.06+382422.2 & 0.137065 & 44.25 & 0.096 & 7.16 & 0.486 \\
SDSS J160344.44+264651.3 & 0.085673 & 44.60 & 0.166 & 7.27 & 0.314 \\
SDSS J160404.51+493820.5 & 0.148582 & 44.00 & 0.092 & 6.93 & 0.186 \\
SDSS J160426.88+525130.3 & 0.107031 & 44.35 & 0.212 & 6.91 & $-$0.686 \\
SDSS J160558.12+440319.5 & 0.044438 & 44.03 & 0.070 & 7.07 & 0.548 \\
SDSS J160806.68+424057.8 & 0.084737 & 44.44 & 0.207 & 7.01 & $-$0.402 \\
SDSS J161527.68+403153.6 & 0.083355 & 44.20 & 0.190 & 6.81 & $-$0.898 \\
SDSS J161713.51+515618.8 & 0.198799 & 44.52 & 0.149 & 7.24 & 0.332 \\
SDSS J161951.31+405847.3 & 0.037858 & 43.77 & 0.062 & 6.87 & 0.350 \\
SDSS J162755.24+470453.0 & 0.271820 & 44.88 & 0.234 & 7.40 & 0.294 \\
SDSS J163128.59+404535.9 & 0.181227 & 44.76 & 0.211 & 7.32 & 0.232 \\
SDSS J163152.22+345328.6 & 0.072234 & 44.28 & 0.107 & 7.14 & 0.400 \\
SDSS J163214.84+333412.8 & 0.174118 & 44.85 & 0.186 & 7.47 & 0.512 \\
SDSS J163247.87+383239.6 & 0.139247 & 44.62 & 0.270 & 7.08 & $-$0.506 \\
SDSS J163417.81+474453.1 & 0.177306 & 44.86 & 0.258 & 7.34 & 0.110 \\
SDSS J163625.42+421346.9 & 0.141250 & 44.70 & 0.308 & 7.10 & $-$0.624 \\
SDSS J163737.38+341205.5 & 0.235658 & 44.56 & 0.150 & 7.27 & 0.384 \\
SDSS J163927.71+354343.3 & 0.317929 & 44.88 & 0.212 & 7.44 & 0.416 \\
SDSS J164100.10+345452.6 & 0.164078 & 44.82 & 0.183 & 7.45 & 0.500 \\
SDSS J164207.32+344834.2 & 0.207533 & 44.60 & 0.144 & 7.33 & 0.486 \\
SDSS J164225.29+391742.2 & 0.184353 & 44.57 & 0.135 & 7.33 & 0.512 \\
SDSS J164416.85+423158.4 & 0.160893 & 44.63 & 0.218 & 7.18 & $-$0.038 \\
SDSS J164626.09+392932.1 & 0.100365 & 44.50 & 0.255 & 6.98 & $-$0.754 \\
SDSS J164907.63+642422.2 & 0.183524 & 44.60 & 0.259 & 7.08 & $-$0.506 \\
SDSS J165437.25+301653.9 & 0.185698 & 44.66 & 0.250 & 7.15 & $-$0.262 \\
SDSS J165636.98+371439.5 & 0.062757 & 43.73 & 0.074 & 6.75 & 0.026 \\
SDSS J165658.36+630051.1 & 0.168969 & 44.55 & 0.253 & 7.04 & $-$0.564 \\
SDSS J165757.51+382327.7 & 0.181496 & 44.74 & 0.191 & 7.35 & 0.350 \\
SDSS J165914.68+313423.4 & 0.264525 & 44.89 & 0.260 & 7.37 & 0.162 \\
SDSS J170002.15+383258.1 & 0.166573 & 44.53 & 0.159 & 7.22 & 0.274 \\
SDSS J170546.91+631059.1 & 0.119182 & 44.49 & 0.141 & 7.23 & 0.368 \\
SDSS J171033.21+584456.8 & 0.280701 & 45.04 & 0.300 & 7.45 & 0.186 \\
SDSS J171526.52+291923.5 & 0.208273 & 44.91 & 0.253 & 7.40 & 0.232 \\
SDSS J171540.93+560654.8 & 0.297139 & 45.12 & 0.246 & 7.62 & 0.548 \\
SDSS J171943.77+581112.3 & 0.350752 & 44.86 & 0.264 & 7.33 & 0.084 \\
\bottomrule
\end{tabularx}
\label{table1_7}
\end{table}

\begin{table}[H]
\caption{All objects with their estimated spin values. Part 8 of 8.}
\footnotesize
\begin{tabularx}{\textwidth}{m{4cm}<{\raggedright}m{1.3cm}<{\centering}m{2.5cm}<{\centering}m{1cm}<{\centering}m{1.3cm}<{\centering}m{1.2cm}<{\centering}}
\toprule
\textbf{Object}	& \textbf{\boldmath$z$} & \textbf{\boldmath$\log(L_\text{bol}\text{[erg/s]})$} & \textbf{\boldmath$l_\text{E}$} & \textbf{\boldmath$M_\text{BH}/M_\odot$} & \textbf{\boldmath$a$}\\
\midrule
SDSS J204404.53-011214.6 & 0.172451 & 44.52 & 0.223 & 7.06 & $-$0.354 \\
SDSS J204731.68+002056.3 & 0.181679 & 44.42 & 0.126 & 7.21 & 0.400 \\
SDSS J205418.80+004915.9 & 0.227509 & 45.16 & 0.243 & 7.66 & 0.598 \\
SDSS J210533.44+002829.3 & 0.054312 & 43.99 & 0.124 & 6.79 & $-$0.402 \\
SDSS J210629.86+110109.0 & 0.304460 & 45.05 & 0.287 & 7.48 & 0.274 \\
SDSS J211436.68-004938.4 & 0.145364 & 44.49 & 0.148 & 7.21 & 0.294 \\
SDSS J212210.99+104200.1 & 0.299016 & 44.95 & 0.232 & 7.48 & 0.400 \\
SDSS J212327.26+001439.9 & 0.182258 & 44.78 & 0.277 & 7.23 & $-$0.180 \\
SDSS J213059.76+004438.0 & 0.130070 & 44.66 & 0.292 & 7.08 & $-$0.624 \\
SDSS J213245.28+121256.8 & 0.125566 & 44.73 & 0.149 & 7.45 & 0.590 \\
SDSS J214054.55+002538.1 & 0.083841 & 44.88 & 0.308 & 7.28 & $-$0.180 \\
SDSS J214249.64-085434.4 & 0.135292 & 44.50 & 0.162 & 7.18 & 0.186 \\
SDSS J214733.86+004021.0 & 0.124192 & 44.64 & 0.277 & 7.09 & $-$0.564 \\
SDSS J215147.60-080922.4 & 0.120572 & 44.68 & 0.204 & 7.26 & 0.162 \\
SDSS J220042.72-073056.4 & 0.110624 & 44.49 & 0.243 & 6.99 & $-$0.624 \\
SDSS J220735.12-082457.7 & 0.213227 & 44.79 & 0.277 & 7.24 & $-$0.142 \\
SDSS J221953.18-083258.7 & 0.305785 & 44.87 & 0.282 & 7.31 & $-$0.004 \\
SDSS J222115.58-004030.4 & 0.171926 & 44.56 & 0.137 & 7.31 & 0.486 \\
SDSS J222255.55+005033.7 & 0.112503 & 44.25 & 0.199 & 6.84 & $-$0.898 \\
SDSS J224605.44-091925.1 & 0.118466 & 44.35 & 0.154 & 7.05 & $-$0.004 \\
SDSS J225452.22+004631.3 & 0.090735 & 44.72 & 0.308 & 7.12 & $-$0.564 \\
SDSS J230108.39-084848.8 & 0.171870 & 44.81 & 0.199 & 7.40 & 0.384 \\
SDSS J230323.47-100235.3 & 0.180613 & 44.90 & 0.285 & 7.34 & 0.026 \\
SDSS J230723.24+001708.1 & 0.112685 & 44.36 & 0.209 & 6.93 & $-$0.624 \\
SDSS J231309.71+002633.7 & 0.285273 & 44.96 & 0.261 & 7.43 & 0.254 \\
SDSS J233811.52+002045.7 & 0.278855 & 44.81 & 0.198 & 7.40 & 0.400 \\
SDSS J234114.22-102828.8 & 0.277839 & 44.96 & 0.219 & 7.51 & 0.474 \\
SDSS J234150.81-004329.0 & 0.250595 & 44.81 & 0.260 & 7.29 & $-$0.004 \\
SDSS J234208.30-094747.5 & 0.191234 & 44.61 & 0.210 & 7.18 & $-$0.038 \\
SDSS J234229.45-004731.4 & 0.315655 & 44.84 & 0.269 & 7.30 & $-$0.004 \\
SDSS J234601.30-101549.0 & 0.191210 & 44.42 & 0.113 & 7.26 & 0.524 \\
SDSS J234725.29-010643.7 & 0.182002 & 44.88 & 0.236 & 7.40 & 0.274 \\
SDSS J235340.46-093709.0 & 0.311753 & 44.86 & 0.227 & 7.39 & 0.294 \\
\bottomrule
\end{tabularx}
\label{table1_8}
\end{table}

\begin{adjustwidth}{-\extralength}{0cm}

\reftitle{References}

\externalbibliography{yes}

\PublishersNote{}
\end{adjustwidth}
\end{document}